\documentclass[journal=jctcce,manuscript=article,layout=twocolumn]{achemso}
\setkeys{acs}{articletitle=true}

\usepackage[version=3]{mhchem} 

\usepackage{lipsum} 

\author{Farren Curtis}
\affiliation{Department of Physics, Carnegie Mellon University, Pittsburgh, PA 15213, USA}
\author{Xiayue Li}
\affiliation{Google, Mountain View, CA 94030, USA}
\author{Timothy Rose}
\affiliation{Department of Materials Science and Engineering, Carnegie Mellon University, Pittsburgh, PA 15213, USA}
\author{\'{A}lvaro V\'{a}zquez-Mayagoitia}
\affiliation{Argonne Leadership Computing Facility, Argonne National Laboratory, Lemont, Illinois, 60439, USA.}
\author{Saswata Bhattacharya}
\affiliation{Department of Physics, Indian Institute of Technology Delhi, Hauz Khas, New Delhi 110016, India}
\author{Luca M. Ghiringhelli}
\affiliation{Fritz-Haber-Institut der Max-Planck-Gesellschaft, Faradayweg 4-6, 14195, Berlin, Germany}
\author{Noa Marom}
\affiliation{Department of Materials Science and Engineering, Carnegie Mellon University, Pittsburgh, PA 15213, USA}
\alsoaffiliation{Department of Physics, Carnegie Mellon University, Pittsburgh, PA 15213, USA}
\alsoaffiliation{Department of Chemistry, Carnegie Mellon University, Pittsburgh, PA 15213, USA}
\email{nmarom@andrew.cmu.edu}

\title[jctctemplate]{GAtor: A First Principles Genetic Algorithm for Molecular Crystal Structure Prediction}

\abbreviations{}
\keywords{key1, key2, key3}

\begin{document}

\begin{tocentry}
\begin{center}\includegraphics{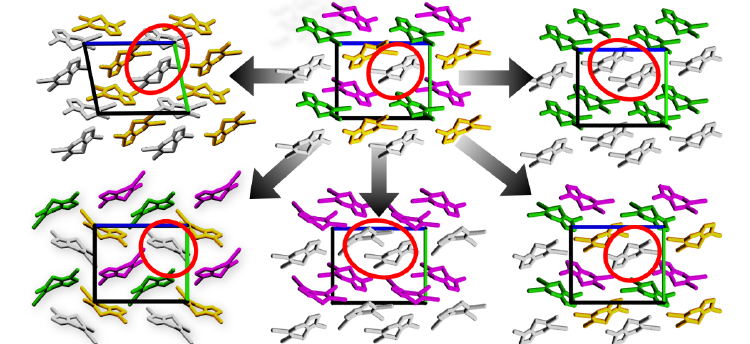}\end{center}
\end{tocentry}


\begin{abstract}
We present the implementation of GAtor, a massively parallel, first principles genetic algorithm (GA) for molecular crystal structure prediction. GAtor is written in Python and currently interfaces with the FHI-aims code to perform local optimizations and energy evaluations using dispersion-inclusive density functional theory (DFT). GAtor offers a variety of fitness evaluation, selection, crossover, and mutation schemes. Breeding operators designed specifically for molecular crystals provide a balance between exploration and exploitation. Evolutionary niching is implemented in GAtor by using machine learning to cluster the dynamically updated population by structural similarity and then employing a cluster-based fitness function. Evolutionary niching promotes uniform sampling of the potential energy surface by evolving several sub-populations, which helps overcome initial pool biases and selection biases (genetic drift). The various settings offered by GAtor increase the likelihood of locating numerous low-energy minima, including those located in disconnected, hard to reach regions of the potential energy landscape. The best structures generated are re-relaxed and re-ranked using a hierarchy of increasingly accurate DFT functionals and dispersion methods. GAtor is applied to a chemically diverse set of four past blind test targets, characterized by different types of intermolecular interactions. The experimentally observed structures and other low-energy structures are found for all four targets. In particular, for Target II, 5-cyano-3-hydroxythiophene, the top ranked putative crystal structure is a $Z^\prime$=2 structure with $P\bar{1}$ symmetry and a scaffold packing motif, which has not been reported previously.
\end{abstract}

\section{Introduction}

Molecular crystals are a unique class of materials with diverse applications in pharmaceuticals, organic electronics, pigments, and explosives \cite{2002Bernstein, 2007Day, 2014Reilly, 2015Elder, 2007Reese, 2009Hasegawa, 2012Bergantin, 2012Cudazzo, 2015Cudazzo,2007Panina, 2015Fitzgerald}. The molecules comprising these crystals are bound by weak dispersion (van der Waals) interactions. As a result, the same molecule may crystallize in several different solid forms, known as polymorphs. Because the structure of a molecular crystal governs its physical properties, polymorphism may drastically impact the desired functionality of a given application. For pharmaceuticals, different polymorphs may display varying stability, solubility, and compressibility, affecting the drug's manufacturability, bioavailability, and efficacy \cite{2013Price, 2016Price, 2002Bernstein}. For applications in organic electronics and organic photovoltaics (OPV), different polymorphs possess different optoelectronic properties\cite{2016Curtis, 2016Wang}, directly impacting device performance \cite{2011Giri,2013Mei,2013Diao}.

\begin{figure}
  \includegraphics{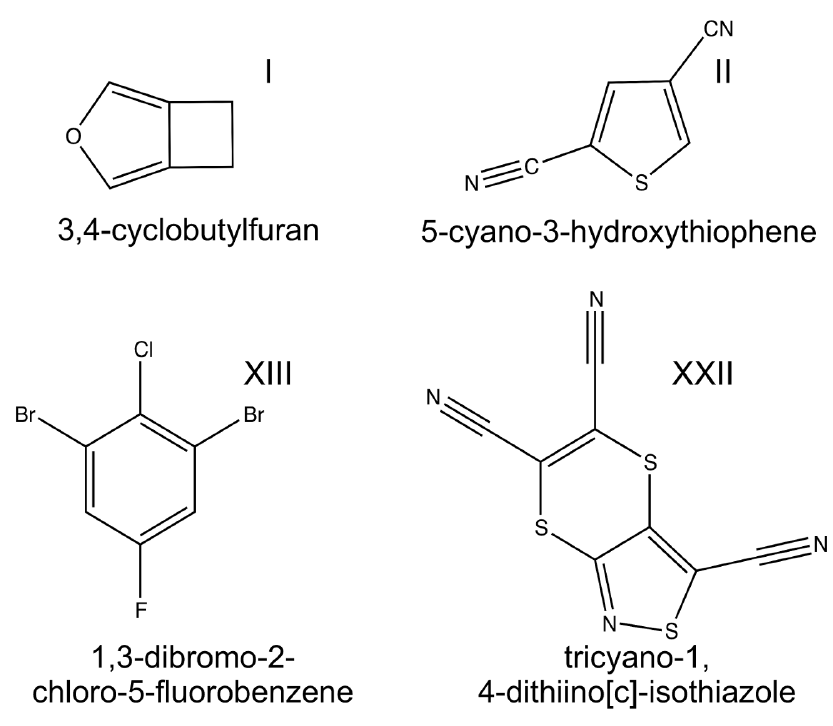}
  \caption{Two-dimensional molecular diagrams of four past blind test targets, Target I\cite{2002Motherwell}, Target II\cite{2002Motherwell}, Target XIII\cite{2009Day}, and Target XXII \cite{2016Reilly}.}
  \label{Fig1_2D_Diagrams}
\end{figure}

Because molecular crystals have a wide range of applications, there has been increasing interest in the fundamental challenge of crystal structure prediction (CSP), or the computation of a molecule's putative crystal structure(s) solely from its two-dimensional chemical diagram, examples of which are shown in Fig. \ref{Fig1_2D_Diagrams}. This challenge is embodied by CSP blind tests, organized periodically by the Cambridge Crystallographic Data Centre \cite{2000Lommerse, 2002Motherwell, 2005Day, 2009Day, 2011Bardwell, 2016Reilly}. CSP can reveal the general behavior of a target molecule, predict the existence of new polymorphs, and serve as a complementary tool for experimental investigations \cite{2015Neumann, 2016Price, 2017Shtukenberg, 2013Meredig}. Once considered unachievable \cite{1994Gavezzotti}, CSP is still an extremely challenging task because it requires combining highly accurate electronic structure methods with efficient algorithms for configuration space exploration.

The energy differences between molecular crystal polymorphs are typically within a few kJ/mol\cite{2013Marom, 2015Cruz, 2015Beran, 2016Beran}, which calls for the accuracy of a quantum mechanical approach. Reaching the required accuracy has become more practical thanks to a decade of development in dispersion-inclusive density functional theory (DFT), including exchange-correlation functionals\cite{2004Dion, 2010Lee, 2009Vydrov, 2011Peverati, 2012Peverati, 2008Zhao, 2010Vydrov, 2010Vydrov_2, 2015Berland, 2015Thonhauser,2016Peng, 2015Sun} and pairwise methods that add the leading order $C^6/R^6$ dispersion term to the inter-nuclear energy\cite{2010Riley, 2006Grimme, 2010Grimme, 2005Johnson, 2012Otero, 2007Jurevcka, 2002Wu, 2001Wu, 2011Steinmann, 2011Steinmann_2,2009Tkatchenko, 2016Brandenburg}. Notably, the recently developed many-body dispersion (MBD) method\cite{2012Distasio, 2012Tkatchenko, 2014Ambrosetti} accurately describes the structure, energetics, dielectric properties, and mechanical properties of molecular crystals\cite{2013Marom, 2013Schatschneider, 2013Reilly, 2013Reilly_2,2014Reilly, 2015Tkatchenko, 2016Curtis, 2017Hermann, 2016Flores} by accounting for long range electrostatic screening and non-pairwise-additive contributions of many-body dispersion interactions.  Using dispersion-inclusive DFT for the final ranking of relative stabilities has become a CSP best practice\cite{2016Reilly}. Vibrational contributions to the zero-point energy and free energy of the system at finite temperature have also been shown to affect the relative stabilities of certain molecular crystal polymorphs and may be further included \cite{2013Reilly_2,2014Reilly, 2017Hoja, 2016Rossi,2015Nyman}.

Approaches to configuration space exploration in CSP include molecular dynamics \cite{2011Yu, 2016Schneider}, Monte Carlo methods \cite{2015Neumann,2013Akkermans}, particle swarm optimization\cite{2012Wang}, and (quasi)-random searches \cite{2011Pickard, 2016Case}. Genetic algorithms (GAs) are a versatile class of optimization algorithms inspired by the evolutionary principle of survival of the fittest \cite{2003Johnston, 2010Sierka, 2013Heiles}. A GA starts from an initial pool of locally optimized trial structures. The scalar descriptor (or combination of descriptors) being optimized is mapped onto a fitness function and structures with higher fitness values are assigned higher probabilities for mating. Breeding operators create offspring structures by combining the structural “genes”\footnote{The term ``genetic algorithm" is sometimes reserved for an evolutionary algorithm that purely encodes an individual's genes with bit-string representations. For our purposes we make no such distinction between genetic and evolutionary algorithms.} of one or more parent structure(s). The child structure is locally optimized and added to the population. The cycle of local optimization, fitness evaluation, and offspring generation propagates structural features associated with the property being optimized and repeats till ``convergence" (a GA is not guaranteed to find the global minimum). For practical purposes, convergence may be defined as when the GA can no longer find any new low-energy structures in a large number of iterations. 

GAs can be applied robustly to complex multidimensional search spaces, including those with many extrema or discontinuous derivatives. They provide a good balance between exploration and exploitation by introducing randomness in the mating step followed by local optimization. Furthermore, they are conceptually simple algorithms, ideal for parallelization, and can lead to unbiased and unintuitive solutions. In the context of structure prediction, the target function being optimized is typically the total or free energy. GAs have been used extensively to find the global minimum structures of crystalline solids\cite{2006Oganov, 2006Glass,2006Abraham,2007Trimarchi, 2011Wu, 1999Woodley, 2008Trimarchi, 2011Lonie, 2002Johannesson, 2012Zhu, 2015Lund,2017Avery,2016Falls} and clusters\cite{1999Morris, 2003Johnston,2005Alexandrova, 2010Sierka,2010Marques,1999Hartke,2010Catlow,2013Heiles,2004Bazterra, 2013Bhattacharya, 2014Bhattacharya,2017Jorgensen, 2013Tipton}. Advantageously, the GA fitness function may be based on any property of interest, not necessarily the energy\cite{2003Johnston, 2011OBoyle, 2013Jain, 2012dAvezac, 2013Zhang, 2010Chua, 2015Bhattacharya}. For organic molecular crystals the goal is not just to locate the most stable structure but also any potential polymorphs. In the most recent CSP blind test \cite{2016Reilly}, GAs were used by us\footnote{In the sixth blind test we used a preliminary version of GAtor.} and others (see submissions \#8, \#12, \#21).

Here, we present GAtor, a new, massively parallel, first principles genetic algorithm (GA) specifically designed for structure prediction of crystal structures of (semi-)rigid molecules. GAtor is written in Python with a modular structure that allows the user to switch between and/or modify core GA routines for specialized purposes. For initial pool generation, GAtor relies on a separate package, Genarris, reported elsewhere \cite{2017Li} and briefly described in Section 3.1. GAtor offers a variety of features that enable the user to customize the search settings as needed for chemically diverse systems, including different fitness, selection, crossover, and mutation schemes. GAtor is designed to fully utilize high performance computing (HPC) architectures by spawning several parallel GA replicas that read from and write to a common population of structures. This approach does not require a full ``generation" of candidates to complete before performing a new selection\cite{2015Bhattacharya, 2013Bhattacharya, 2014Bhattacharya}. For energy evaluations and local optimization of trial structures, GAtor employs dispersion-inclusive DFT by interfacing with the ab initio, all-electron electronic structure code FHI-aims \cite{2009Blum,2009Havu}. 

The paper is organized as follows: Section 2 describes the DFT methods and numerical settings of FHI-aims used in conjunction with GAtor; Section 3 details GAtor's parallelization scheme and the features currently available in the code; Section 4 showcases applications of GAtor for a chemically diverse set of four past blind test targets, 3,4-cyclobutylfuran (Target I\cite{2002Motherwell}), 5-cyano-3-hydroxythiophene (Target II\cite{2002Motherwell}), 1,3-dibromo-2-chloro-5-fluorobenzene (Target XIII\cite{2009Day}), and tricyano-1,4-dithiino[c]-isothiazole (Target XXII\cite{2016Reilly}) shown in Fig. \ref{Fig1_2D_Diagrams}. Finally, Section 5 provides concluding remarks and best practices.

\section{DFT Settings}
Because first principles calculations are computationally expensive, lighter DFT settings are employed within the GA search, with the intention of locating the experimental structure and any potential polymorphs among the lowest energy structures. To obtain more precise rankings, the best structures produced from the GA are postprocessed with higher-level functionals and dispersion corrections. Hierarchal screening approaches have become a common practice in CSP \cite{2016Reilly}. In GAtor, the user has the option to input FHI-aims control files for any desired level(s) of theory. The DFT settings used in the present study are detailed below.

For local structural optimizations within the GA, the generalized gradient approximation of Perdew-Burke-Ernzerhof (PBE)\cite{1996Perdew, 1997Perdew} is used with the pairwise Tkatchenko-Scheffler (TS) dispersion-correction\cite{2009Tkatchenko} with \textit{lower-level} numerical settings, which correspond to the light numerical settings and tier 1 basis sets of FHI-aims \cite{2009Blum}. During local optimization, the space group symmetry is allowed to vary. Additionally, a $2 \times 2 \times 2$ k-point grid and reduced angular grids are used.  A convergence value of $10^{-5}$ electrons is set for the change in charge density in the self-consistent field (SCF) cycle and SCF forces and stress evaluations are not computed. These settings are implemented in order to accelerate geometry relaxations within the GA.  For Target XIII, atomic ZORA scalar relativity\cite{2009Blum} settings are used for the heavier halogen elements.

For postprocessing, the best 5-10\% of the final structures produced by the GA are re-relaxed and re-ranked using a $3 \times 3 \times 3$ k-point grid, PBE+TS, and \textit{higher-level} numerical settings, which correspond to the tight/tier2 default settings of FHI-aims\cite{2009Blum}. Next, single point energy (SPE) evaluations are performed using PBE with the MBD method \cite{2012Distasio, 2012Tkatchenko, 2014Ambrosetti} for the best structures as ranked by PBE+TS. The final re-ranking is performed using the hybrid functional PBE0 \cite{1996Perdew_2, 1999Adamo} with the MBD correction. The inclusion of 25\% exact exchange in PBE0 mitigates the self-interaction error, leading to a more accurate description of electron densities and multipoles \cite{2013Reilly, 2013Reilly_2}. For some molecular crystals the correct polymorph ranking is reproduced only when using PBE0+MBD \cite{2016Curtis, 2013Marom}. The PBE0+MBD ranking is considered to be the most reliable of the methods used here. Thermal contributions to the total energy, shown to change the energy ranking in approximately 9\% of organic compounds\cite{2015Nyman}, are not further included in the present study.

\section{Code Description}
GAtor is written in Python and uses the spglib \cite{2017Spglib} crystal symmetries library, sci-kit learn\cite{2011scikitlearn} machine learning package, and pymatgen\cite{2013Ong} library for materials analysis. GAtor is available for download from www.noamarom.com under a BSD-3 license. The code is modular by design, such that core GA tasks, such as selection, similarity checks, crossover, and mutation can be interchanged in the user input file and/or modified. For energy evaluations and local optimization GAtor currently interfaces with the all-electron DFT code FHI-aims  \cite{2009Blum,2009Havu}, and may be modified to interface with other electronic structure and molecular dynamics packages.

\begin{figure}[h!]
  \includegraphics{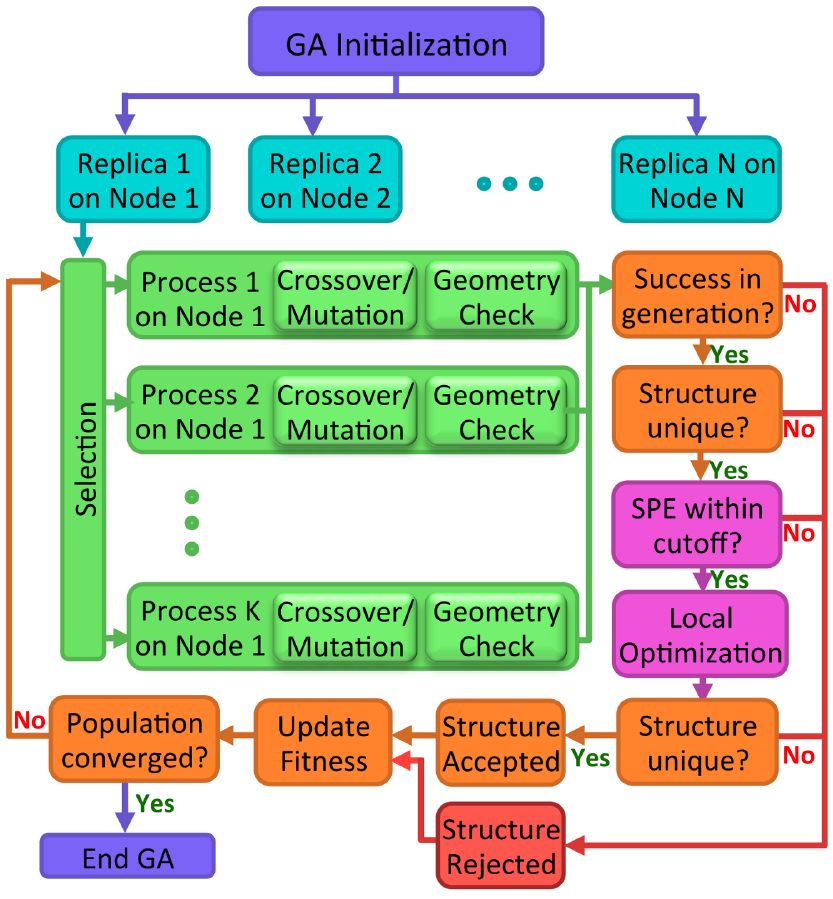}
  \caption{An example workflow of GAtor on a high performance computing cluster. In the diagram, $N$ independent GA replicas run on $N$ computing nodes, with $K$ core processing units per node. Single point energy (SPE) evaluations and local optimizations are performed using FHI-aims.}
  \label{Fig2_GAtor_parallelization}
\end{figure}
GAtor takes advantage of high performance computing (HPC) architectures by avoiding processor idle time and effectively utilizing all available resources. An example workflow is shown in Fig. \ref{Fig2_GAtor_parallelization}. After initialization, the master process spawns a user-defined number of GA replicas across $N$ nodes. Each independent replica performs the core genetic algorithm tasks independently while reading from and writing to a dynamically updated pool of structures\cite{2013Bhattacharya,2014Bhattacharya,2015Bhattacharya}. Additional multiprocessing may be utilized within each replica for child generation. GAtor has been tested on up to 16,384 Blue Gene/Q nodes (262,144 cores) at the Argonne Leadership Computing Facility. 

Two classes of breeding operators are implemented in GAtor, crossover and mutation, described in detail in Sections 3.4-3.5. Crossover operators generate a child by combining the structural genes of two parents, whereas mutation operators create a child by altering the structural genes of one parent. After selection, either crossover or mutation is performed with a user-defined probability. When multiprocessing is used, the same set of parents (crossover) or single parent (mutation) undergo the same breeding operation, but with different random parameters. If a child cannot pass the geometry checks after a user-defined number of attempts, a new selection is performed. Otherwise, the first child that passes the geometry checks proceeds to the first uniqueness check.  If a candidate structure successfully passes all geometry checks, uniqueness checks, and energy cutoffs, it is added to the common population. The fitness of each structure in the population is updated, and a new selection can be performed immediately. A detailed account of the core tasks and features of the GA is provided below.

\subsection{GA Initialization}
During GA initialization GAtor reads in an initial pool of structures generated by the Genarris random structure generation package\cite{2017Li} using the diverse workflow. Genarris generates random symmetric crystal structures in the 230 crystallographic space groups and then combines fragment-based DFT with clustering techniques from machine learning to produce a high-quality, diverse starting population at a relatively low computational cost, as described in detail in Ref. \citenum{2017Li}. The initial pool structures are pre-relaxed with PBE+TS and \textit{lower-level} numerical settings as described in Section 2 and their total energies are stored beforehand. GAtor updates the starting fitness values of the initial pool structures, as described below, before performing selection.

\subsection{Fitness Evaluation}
The fitness of an individual determines its likelihood of being chosen for crossover or mutation. GAtor provides a traditional energy-based fitness function, in which structures with lower relative stabilities are assigned higher fitness values. Additionally, GAtor provides the option of a cluster-based fitness function, which can use various clustering techniques to perform evolutionary niching. Using cluster-based fitness can reduce genetic drift, as explained below, by suppressing the over-sampling of certain regions of the potential energy surface and promoting the evolution of several subpopulations simultaneously. 

\subsubsection{Energy-based Fitness}
In energy-based fitness, the total energy $E_i$ of the $i$th structure in the population is evaluated using dispersion-inclusive DFT as detailed in Section 2. The fitness $f_i$ of each structure is defined as,
\begin{align}
f_i&= \frac{\epsilon_i}{\sum_{i}\epsilon_i}\hspace{1cm}0\leq f\leq 1\\
\epsilon_i&=\frac{E_{\mbox{\footnotesize{max}}}-E_{{\footnotesize{i}}}}{E_{\mbox{\footnotesize{max}}}-E_{\mbox{\footnotesize{min}}}}
\end{align}
where $\epsilon_i$ is the $i$th structure's relative energy, and  $E_{\footnotesize{\mbox{max}}}$ and $E_{\footnotesize{\mbox{min}}}$ correspond to the structures with the dynamically updated highest and lowest total energies in the population, respectively\cite{2015Bhattacharya, 2013Bhattacharya, 2014Bhattacharya}. Hence, structures with lower relative energies have higher fitness values.

\subsubsection{Cluster-Based Fitness}
When using a traditional energy-based fitness function, a GA may be prone to exploring the same region(s) of the potential energy surface, which may or may not include the experimental structure(s) or the global minimum structure. This may be due to a number of factors, including lack of diversity in the common population and biases towards or against certain packing motifs over time, a phenomenon known as genetic drift.  Genetic drift can result from  biases in the initial pool\cite{2017Li} and from the topology of the potential energy landscape (e.g. a desirable packing motif for a given molecule could be located in narrow well that is rarely visited). The search may also be influenced by systematic biases of the energy method used (e.g., the exchange-correlation functional and dispersion method), towards or against certain packing motifs\cite{2016Curtis}.

GAs may be adapted to be more suitable for multi-modal optimization using evolutionary niching methods\cite{1998Sareni, 2012Shir, 2015Preuss}. Niching methods support the formation of stable subpopulations in the neighborhood of several optimal solutions. For molecular crystal structure prediction, incorporating niching techniques may increase diversity and diminish the effect of inherent or initial pool biases. The goal is for the GA to locate all low-energy polymorphs that may or may not have similar structural motifs to the experimentally observed crystal structure(s) or the most stable crystal structure present in the population. 

GAtor provides the option to dynamically cluster the common population of molecular crystals into groups (niches) of structural similarity, using pre-defined feature vectors for each target molecule and clustering algorithms implemented in the sci-kit learn machine learning Python package\cite{2011scikitlearn}. Currently, GAtor offers the use of radial distribution function (RDF) vectors of interatomic distances for user-defined species, relative coordinate descriptor (RCD) vectors \cite{2017Li}, or a simple lattice parameter based descriptor, $L$, given by:
\begin{equation}
L=\frac{1}{\sqrt[3]{V}}(a, b, c)
\end{equation}
where $V$ is the unit cell volume and $a$, $b$, and $c$ are the structure's lattice parameters after employing Niggli reduction\cite{1928Niggli, 1973Gruber, 1976Kvrivy, 2004Grosse} and unit cell standardization. Niggli reduction produces a unique representation of the translation vectors of the unit cell but does not define a standard orientation. Therefore, all unit cell lattice vectors are standardized such that $\vec{a}$ points along the $\hat{x}$ direction, $\vec{b}$ lies in the $xy$ plane, and the convention $a\leq b\leq c$ is used. The lattice parameter based descriptor encourages the sampling of under-represented lattices in the population (e.g. structures which are almost 2D which may have one lattice parameter significantly shorter than the others). GAtor offers K-Means\cite{2002Kanungo} and Affinity Propagation (AP)\cite{2007Frey} clustering, and may be adapted to use other clustering algorithms implemented in sci-kit learn. AP is a clustering method that determines the number of clusters in a data set, based on a structure similarity matrix, rather than defining the number of clusters \textit{a priori}. This has the advantage of resolving small, structurally distinct clusters\cite{2017Li}. Once the common population has been clustered into niches, a fitness sharing scheme\cite{1998Sareni} is applied such that a structure's scaled fitness, $f'_i$, is given by
\begin{equation}
f'_i = \frac{f_i}{ m_i}
\end{equation}
where $m_i$ is a cluster-based scaling parameter, currently determined by the number of structures in each individual's shared cluster. This clustering scheme increases the fitness of under-sampled low-energy motifs within the population, and suppresses the over-sampling of densely populated regions. One example of evolutionary niching is discussed in Section 4.1 for Target XXII. Further investigations of the effect of the descriptor and the fitness function will be the subject of future work.

There are a variety of other strategies for incorporating niching or clustering into an evolutionary algorithm. Refs. \citenum{2010Lyakhov}-\citenum{2013Lyakhov} use fingerprint functions based on inter-atomic distances to prevent too dissimilar structures from mating. Recently, Ref. \citenum{2017Jorgensen} explored incorporating agglomerative hierarchical clustering (AHC) into an evolutionary algorithm applied to organic molecules and surfaces. AHC detects the number of clusters in the given data set, similar to AP. One of their methods promoted selection of cluster outliers, while another utilized a fitness function that combined the structure's cluster size with its energy, similar to the technique employed in GAtor.

\subsection{Selection}
Selection is inspired by the evolutionary principle of survival of the fittest. In GAtor, individuals with structural motifs associated with higher fitness values have a higher probability of being selected for mating. GAtor currently offers a choice of two genetic algorithm selection strategies: roulette wheel selection and tournament selection. 

\subsubsection{Roulette wheel selection}
This selection technique \cite{1989Goldberg} simulates a roulette wheel, where fitter individuals in the population conceptually take up larger slots on the wheel, and therefore have a higher probability of being selected when the wheel is spun. In GAtor, the procedure is as follows: First, a random number $r$ is chosen, uniform in the interval [0, 1]. Then, a parent structure is selected for mating if it has the first sorted, normalized fitness value with  $f_i>r$\cite{2013Bhattacharya, 2014Bhattacharya,2015Bhattacharya}.

\subsubsection{Tournament Selection}
In tournament selection \cite{1989Goldberg}, a user-defined number of individuals are randomly selected from the common population to form a tournament. In GAtor, the two structures with the highest fitness values in the tournament (i.e. the winner and the runner-up) are selected for mating. Tournament selection is efficient (requiring no sorting of the population) and gives the user control over the selection pressure via control of the tournament size \cite{1996Blickle}.


\subsection{Crossover}
Crossover is a breeding operator that combines the structural genes of two parent structures selected for mating to form a single offspring.  The crossover operators implemented in GAtor were developed specifically for organic molecular crystals. The popular `cut-and-splice'\cite{1995Deaven} crossover operator used in other genetic algorithms, takes a random fraction of the each parent's unit cell (and the motifs within) and pastes them together.  While this approach is successful for structure prediction of clusters and inorganic crystals\cite{2004Bazterra, 1999Hartke, 2006Oganov, 2006Glass,2008Trimarchi,2009Froltsov, 2010Ji,2011Lonie,2013Bhattacharya, 2014Bhattacharya,2015Bhattacharya,2017Jorgensen}, it may not be the most natural choice for molecular crystals because it can break important space group symmetries that may be associated with, e.g., efficient packing and lower total energies. Initialization of the starting population within random symmetric space groups has been shown to increase the efficiency of evolutionary searches \cite{2010Wang,2012Zhu,2013Lyakhov, 2017Avery_2}. In the same vein, further steps can be taken to design the breeding operators themselves to exploit and explore the symmetry of the starting population and to reduce the number of expensive first principles calculations on structures far from equilibrium. Therefore, several mutation and crossover operators implemented in GAtor can preserve or break certain space group symmetries of the parent structure(s), as detailed below.

\subsubsection{Standard Crossover}
In this crossover scheme each parent's genes are represented by the Niggli-reduced, standardized unit cell lattice parameters and angles ($a, b, c, \alpha, \beta, \gamma$) as well as the molecular geometry\footnote{The geometry of the molecules are allowed to relax during local optimization. This is important for semi-rigid molecules, such as Target XXII. This extra degree of freedom is accounted for in the crossover process by randomly selecting the relaxed molecular geometry from one parent.}, orientation $\Phi=(\theta_z, \theta_y, \theta_x)$, and center of mass (COM) position in fractional coordinates, $R_{\mbox{\tiny{COM}}}$, of each molecule within the unit cell. The orientation of each molecule within the unit cell is defined by computing the $\theta_z$, $\theta_y$, and $\theta_x$ Euler angles, respectively, which rotate a Cartesian reference frame to an inertial reference frame aligned with each molecule's principal axes of rotation. When generating a child structure, the molecules in the unit cell of each parent structure are randomly paired together. The fractional COM positions for each molecule in the child structure are directly inherited from one randomly selected parent. The lattice parameters from each parent are combined with random fractions to form the lattice parameters of the child structure. The child's molecular geometries are inherited from one randomly selected parent and initially centered at the origin with their principal axes of rotation aligned with the Cartesian axes. The final orientations of the molecules in the child structure are constructed by combining the orientation angles of the paired molecules from the parent structures with random fractions.

\begin{figure}
  \includegraphics{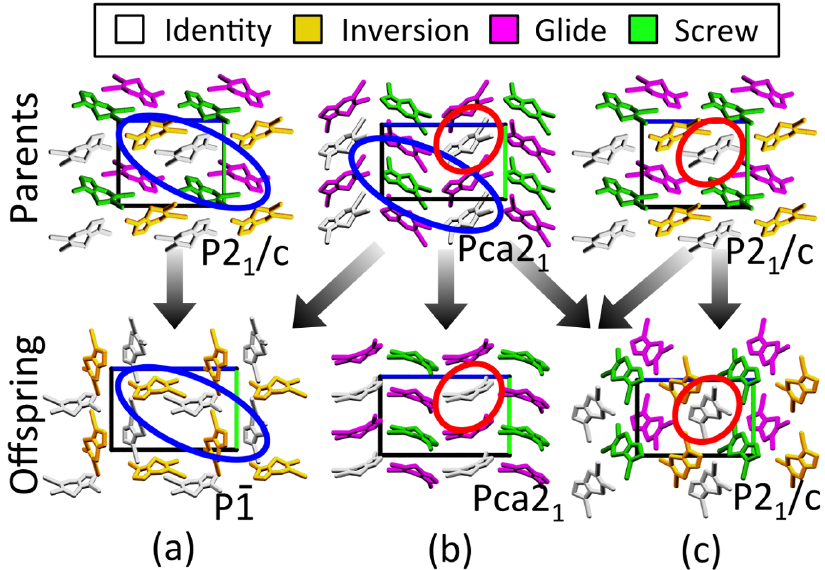}
  \caption{Examples of (a) standard crossover and (b-c) symmetric crossover applied to selected parent structures of Target XXII. The colors of the molecules correspond to the symmetry operations applied to the asymmetric unit of each structure, shown in white. The structures shown are projected along the $\vec{a}$ lattice vector and the $\vec{b}$, and $\vec{c}$ lattice vectors are highlighted in green and blue, respectively.}
  \label{Fig3_Crossover_methods}
\end{figure}

Fig. \ref{Fig3_Crossover_methods}, panel (a) shows an example of standard crossover for two selected parent structures of Target XXII with space groups $P2_1/c$ and $Pca2_1$, respectively. Four molecules from each parent are randomly selected (circled in blue) and paired together. The molecular geometries and COM positions of the child structure are both inherited from the $P2_1/c$ parent structure. The orientation angles of the molecules paired from each parent structure are combined with random fractions. The lattice parameters are also combined with random fractions. In this specific example, a child structure is created with a $Z^\prime=2$ motif that has lower symmetry than either of its parents, $P\bar{1}$, but still contains inversion symmetry before local optimization.

\subsubsection{Symmetric Crossover}
In this crossover scheme each parent's genes are represented by the orientation and COM position of their respective crystallographic asymmetric units as well as their respective space group operations and unit cell lattice parameters. For the explicit computation of each parent's asymmetric unit and space group operations, GAtor relies on the pymatgen\cite{2013Ong} package, which utilizes the spglib crystal symmetries library\cite{2017Spglib}. When generating a child structure, the genes of the parents are combined strategically to preserve one parent's space group as detailed below. 

First, the asymmetric unit and corresponding space group operations are deduced for both parents. If the two asymmetric units contain the same number of molecules, then the respective molecules in each unit are paired together. If the asymmetric units contain a different number of molecules, then one parent's asymmetric unit is used as a reference and paired with an equivalent number of molecules in the second parent's unit cell. If the asymmetric units contain different relaxed molecular geometries, then the molecular conformations in the child's asymmetric unit may be randomly inherited from one parent. The orientation and COM position of the molecule(s) within the child's asymmetric unit are constructed by combining the orientation and COM position of the paired molecule(s) from each parent with random fractions. If both parents possess the same Bravais lattice type then their lattice parameters may be combined with random fractions. Otherwise, the child's lattice is randomly inherited from one parent.  Finally, the symmetry operations (containing specific translations, reflections, and rotations of the asymmetric unit in fractional coordinates) are selected from one parent and applied to the child's generated asymmetric unit and lattice. Either parent's space group operations may be randomly selected and applied to the child's asymmetric unit when both parents possess the same number of molecule's in the asymmetric unit and the same Bravais lattice type. Otherwise, one parent's space group operations will be compatible with the symmetry of the generated lattice and asymmetric unit by construction and are thus applied. This crossover procedure ensures the space group of the child is directly inherited from one of its parents, at least before local optimization, which does not constrain the symmetry of the child structure.

Examples of symmetric crossover are shown in Fig. \ref{Fig3_Crossover_methods}, panels (b) and (c).  The participating asymmetric units of the parent and child structures are circled in red. In panel (b), the child structure inherits the molecular geometry from the $Pca2_1$ parent structure, which is more planar than the molecular geometry of the $P2_1/c$ structure. The orientations of the asymmetric units (both $Z^\prime$=1) and lattice vectors of both parents are combined with random weights. The space group symmetry operations from the $Pca2_1$ parent are applied to the child's asymmetric unit on the generated lattice. In panel (c), the child structure inherits the molecular geometry and symmetry operations from the $P2_1/c$ parent structure. The randomness used when creating the orientation of the motif in the asymmetric unit explains why the child shown in panel (b) has a different orientation of the asymmetric unit as the one shown in panel (c), and allows for more diversity in the generated offspring. In these specific examples, both child structures produced using symmetric crossover have higher symmetry than the child produced with standard crossover, before local optimization.

\subsection{Mutation}
Mutation operators are applied to the genes of single parent structures to form new offspring. In GAtor, certain mutations may promote exploration of the potential energy surface via dramatic structural changes, while others may exploit promising regions via subtle changes. The user chooses the percentage of selected structures that undergo mutation, and may select specific or random mutations to be applied. GAtor also provides an option that allows a percentage of structures to undergo a combination of any two mutation operations before local optimization. This approach encourages exploration and may reduce the number of duplicate structures generated in the search \cite{2011Lonie}.

\subsubsection{Strains} 
\begin{figure}[h!]
  \includegraphics{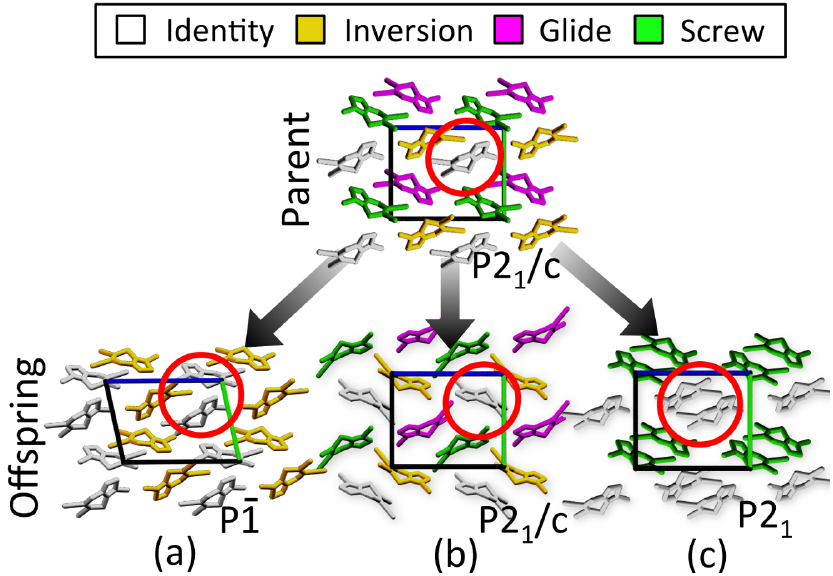}
  \caption{Examples of (a) random strain, (b) rotation, and (c) translation  mutations applied to a $P2_1/c$ structure of Target XXII. The colors of the molecules correspond to the symmetry operations applied to the asymmetric unit of each structure, shown in white and circled in red.  The structures shown are projected along the $\vec{a}$ lattice vector and the $\vec{b}$, and $\vec{c}$ lattice vectors are highlighted in green and blue, respectively.}
  \label{Fig4_Mutation_methods}
\end{figure}

GAtor offers a variety of strain operators that produce child structures by acting upon the lattice vectors of the selected parent structure. Similar to Refs. \citenum{2006Oganov},\citenum{2012Zhu}, and \citenum{2011Lonie}, the strain tensor is represented using the symmetric Voigt strain matrix $\boldsymbol{\epsilon}$,
\begin{equation} 
\boldsymbol{\epsilon} =
   \begin{bmatrix}
    \epsilon_{11} & \frac{\epsilon_{12}}{2} & \frac{\epsilon_{13}}{2} \\
  \frac{\epsilon_{12}}{2} &  \epsilon_{22} & \frac{\epsilon_{23}}{2}\\
    \frac{\epsilon_{13}}{2} & \frac{\epsilon_{23}}{2} & \epsilon_{33}\\
    \end{bmatrix}.
\end{equation}
The strain matrix is applied to each lattice vector $\vec{a}_{\mbox{\tiny parent}}$ of the chosen parent structure to produce the lattice vector of the child $\vec{a}_{\mbox{\tiny child}}$ via 
\begin{equation}
\vec{a}_{\mbox{\tiny child}} =\vec{a}_{\mbox{\tiny parent}} + 
\boldsymbol{\epsilon} \vec{a}_{\mbox{\tiny parent}}.
\end{equation}
The components of $\epsilon_{ij}$ are chosen to produce different modes of strain. To apply random strains, all six unique $\epsilon_{ij}$ components are randomly selected from a normal distribution with a user-defined standard deviation that determines the strength of the applied strain. To apply random deformations in certain crystallographic directions, one or more random $\epsilon_{ij}$ may be chosen while the others are set to 0. Strains that preserve the overall unit cell volume of the parent structure, or change a single unit cell angle, may also be applied. When applying a strain, the COM of each molecule is moved according to its fractional coordinates. An example strain mutation is shown in Fig. \ref{Fig4_Mutation_methods}, panel (a).  Here, a random strain is applied that transforms the lattice of the parent structure from monoclinic ($\alpha=\gamma=90; \beta \neq 90$) to triclinic ($\alpha \neq \beta \neq \gamma \neq 90$). The COM of each molecule is moved accordingly, breaking the glide and screw symmetry of the parent structure and creating a $Z^\prime$=2 child structure. 

\subsubsection{Molecular Rotations}
Rotation mutations change the orientations of the molecules in the selected parent structure. Different random rotations may be applied to the Cartesian coordinates of the atoms in selected molecules centered at the origin, or the same random rotation can be applied about each molecule's principal axes of rotation. For $Z^\prime$=1 structures, the latter type of rotation is equivalent to randomly changing the orientation of molecule in the asymmetric unit, as shown in Fig. \ref{Fig4_Mutation_methods}, panel (b). Here, each molecule from the parent structure receives the same random rotation about its principal axes of rotation, rotating the asymmetric unit and preserving the parent's $P2_1/c$ symmetry in the resulting offspring.

\subsubsection{Translations}
Translational mutations change the position of $R_{\mbox{\tiny{COM}}}$ for certain molecules within the unit cell. They are either applied randomly to the COM (in Cartesian coordinates) of randomly selected molecules, or in a random direction in the basis of the each molecule's inertial reference frame, constructed from each molecule's principal axes of rotation. An example of the latter type of mutation is depicted in Fig. \ref{Fig4_Mutation_methods}, panel (c). Here, each molecule from the parent structure receives the same random translation in the basis of its inertial reference frame. In this case, paired enantiomers are translated in equal and opposite directions, which breaks the glide symmetry of the parent structure, and forms an asymmetric unit containing two molecules in a tightly packed dimer.

\subsubsection{Permutations}
\begin{figure}[h]
  \includegraphics{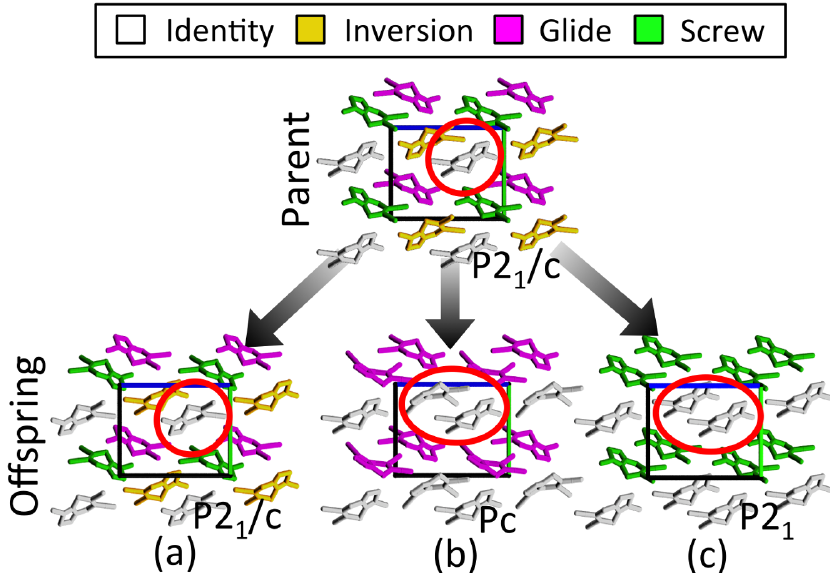}
  \caption{Examples of (a) permutation, (b) permutation-rotation, and (c) permutation-reflection mutations applied to a $P2_1/c$ structure of Target XXII. The colors of the molecules correspond to the symmetry operations applied to the asymmetric unit of each structure, shown in white and circled in red. The structures shown are projected along the $\vec{a}$ lattice vector and the $\vec{b}$, and $\vec{c}$ lattice vectors are highlighted in green and blue, respectively.}
  \label{Fig5_Mutation_methods_2}
\end{figure}

Permutation mutations swap $R_{\mbox{\tiny{COM}}}$ for randomly selected molecules in the parent unit cell. Depending on the point group symmetry of the molecule, the lattice, and the permutation, this operator can preserve, add, or break certain space group symmetries of the parent structure. An example permutation mutation that preserves the parent's space group symmetry is shown in Fig. \ref{Fig5_Mutation_methods_2}, panel (a). Here, a permutation is applied which effectively swaps $R_{\mbox{\tiny{COM}}}$ of the highlighted asymmetric unit (shown in white) and its nearest neighbor (shown in yellow), as well as swapping $R_{\mbox{\tiny{COM}}}$ of the two other molecules in the unit cell related by screw and glide symmetry (shown in green and fuchsia, respectively). As a result, the child structure inherits the $P2_1/c$ symmetry of the parent structure.

\subsubsection{Permutation-Rotations and Permutation-Reflections} 
Permutation-rotation mutations swap randomly selected molecules within the unit cell and then apply a random rotation about their principal axes of rotation. Fig. \ref{Fig5_Mutation_methods_2}, panel (b) shows an example of permutation-rotation. Here, the two molecules in the parent unit cell colored in yellow and green swap position and undergo a random rotation, while the others remain fixed. As a result, the structure produced (space group $Pc$) no longer contains the exact two fold screw symmetry of the parent structure (space group $P2_1/c$) and effectively contains an asymmetric unit consisting of two molecules with the same chirality. In the permutation-reflection mutation, half of the molecules in the unit cell swap positions and then undergo a reflection in the $xy$, $yz$, or $zx$ Cartesian planes centered at their COM. Fig. \ref{Fig5_Mutation_methods_2}, panel (c) shows an example of permutation-reflection. Here, the two molecules in the parent unit cell colored in yellow and fuchsia swap positions and undergo a reflection about the \textit{zx} plane pointing out of the page, while the others remain fixed. As a result, the structure produced (space group $P2_1$) no longer contains the glide symmetry of the parent structure (space group $P2_1/c$), and effectively contains an asymmetric unit consisting of two molecules of the same chirality. For crystals containing chiral molecules, such as Target XXII, this mutation can be especially effective because it can swap the relative positioning of enantiomers within the unit cell.

\subsection{Rejection Criteria}
Because crossover and mutation operations are performed randomly on a diverse set of structures, the offspring generated may be unphysical or duplicates of existing structures. GAtor applies various criteria for rejecting a child structure before performing local optimization.  This preserves the diversity of the population by preventing uncontrolled multiplication of similar structures and avoids computationally expensive local optimization of unreasonable or redundant structures.

\subsubsection{Geometry Checks}
Structures may be rejected if any two intermolecular contacts are too close.  The minimum distance $d{\mbox{\tiny min}}$ between any two atoms $A$ and $B$ belonging to different molecules is given by:
\begin{equation}
d{\mbox{\tiny min}} = s_{\mbox{\footnotesize r}}(r_{\mbox{\tiny A}} +r_{\mbox{\tiny B}})
\end{equation}
where $r_A$ and $r_B$ are the vdW radii of the atoms $A$ and $B$, respectively, and $s_{\mbox{\footnotesize r}}$ is a user-defined parameter typically set between 0.6-0.9.  Additionally, the user may constrain how close the COMs of any two molecules are allowed to be, or specify the allowed unit cell volume range for the generated structures. If the children produced by a parent or set of parents do not pass the geometry checks after a user-defined number of attempts, a new selection is performed.

\subsubsection{Similarity Checks}
Identifying duplicate crystal structures is critical for maintaining diversity and preventing a GA from getting stuck in a specific region of the potential energy surface. Furthermore, it is imperative to identify structures that are too similar to others in the existing population before local optimization to avoid expensive and redundant DFT calculations. Checking for duplicates is complicated by the fact that multiple representations exist for the same crystal structure. To address this issue,  Niggli reduction\cite{1928Niggli, 1973Gruber, 1976Kvrivy, 2004Grosse} and cell standardization are used for all structures within GAtor, as previously described in Section 3.2.2.

GAtor performs a similarity check on all generated offspring before and after local optimization. The pre-relaxation similarity check prevents the local optimization of any structures too similar to others in the population, using loose site and lattice parameter tolerances in pymatgen's StructureMatcher class\cite{2013Ong}. The post-relaxation similarity check identifies whether any optimized structures relaxed into \textit{bona fide} duplicates of existing structures in the population, using stricter site and lattice parameter tolerance settings. If the candidate structure is found to have a similar lattice to another in the common pool (within the user-defined tolerances for the lattice parameter lengths and angles), then the root mean square (RMS) distances are computed between equivalent atomic sites. If the maximum, normalized RMS distance is within the user-defined tolerance, then the two structures are determined to be duplicates.

\subsubsection{Single Point Energy (SPE) Cutoff}
Single point DFT calculations, using PBE+TS and \textit{lower-level} numerical settings, are performed on unrelaxed offspring to decide whether they should undergo local optimization, as shown in Fig. \ref{Fig2_GAtor_parallelization}. 
If the energy of the unrelaxed structure is higher than the user-defined cutoff, it is immediately rejected. This reserves computational resources for the local optimization of structures with energies that are more likely to have desirable genetic features. The energy cutoff can be fixed or set relative to the current global minimum.  Typically, the relative energy cutoff is set to 70-120 kJ/mol per molecule, however it may be system dependent. A recommended best practice is to set the cutoff to prevent the addition of structures worse in energy than those in the diverse initial pool. 

\subsection{Termination}
Because there is no unique way of converging a genetic algorithm, the user specifies simple conditions for when the code should terminate.  One option is choosing to terminate the algorithm if a certain number of the best structures in the common population have not changed in a user-defined amount of iterations (e.g. if the top 20 structures have not changed in 50 iterations of the GA). This tracks whether all low-energy structures have been located in a reasonable number of iterations. Here, an iteration is defined as when a structure has passed all rejection criteria and is added to the common pool. Alternatively, the user may choose to terminate after the total population has reached a certain size. Additionally, the user may terminate the code manually at any time. If GAtor stops due to, e.g. wall time limits or hardware failures, there is an option to restart the code and finish all calculations leftover from the previous run before performing new selection. Code restarts can also be used strategically to modify the GA settings (e.g. to tighten the energy cutoffs or change mutation schemes) without affecting
the common population of structures.
\section{Applications}
GAtor was used to perform crystal structure prediction for the four chemically diverse blind test targets shown in Fig. \ref{Fig1_2D_Diagrams}. The initial pool for each target was generated with Genarris \cite{2017Li} to create a starting population of diverse, high-quality structures\cite{2017Li}. The distribution of space groups for each initial pool is provided in the supporting information. The generated initial pool structures were locally optimized with the same DFT settings used in the GA and checked for duplicates. For each molecule, a variety of crossover, mutation, and selection parameters were run on the same initial population. For testing purposes, GA searches were performed only with the same number of molecules per unit cell as the experimental structure(s). The number of molecules in the asymmetric unit was not constrained. In all cases, the experimental structures were generated as well as several other low-energy structures that may be viable polymorphs. 

\subsection{Target XXII}

Target XXII (C$_8$S$_3$N$_4$) was selected from the sixth blind test\cite{2016Reilly}. It belongs to a unique class of compounds, called thiacyanocarbons, which only contain carbon, nitrogen, sulfur and a plurality of cyano groups\cite{1962Simmons}. 
\begin{figure}[h]
  \includegraphics{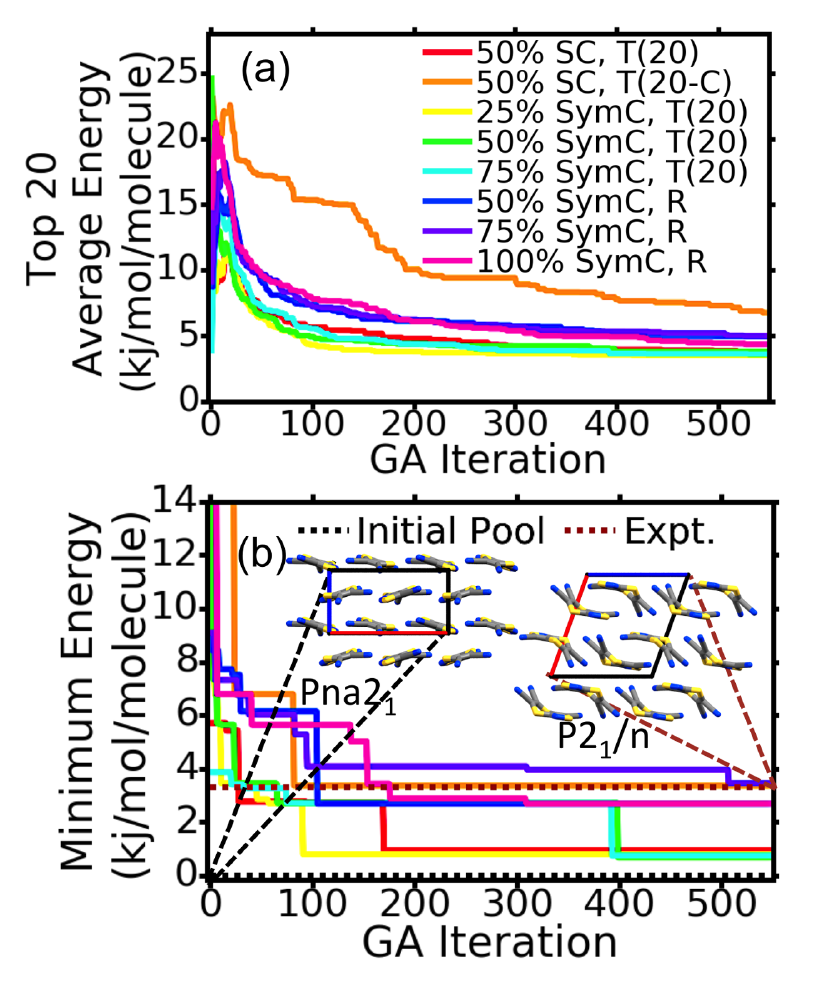}
   \caption{(a) The average energy of the top 20 Target XXII structures as a function of GA iteration and (b) the global minimum structure generated as a function of GA iteration, shown for different GA runs. S, N, and C atoms are colored in yellow, blue, and grey, respectively.  The structures shown are projected along the $\vec{b}$ lattice vector and the $\vec{a}$, and $\vec{c}$ lattice vectors are highlighted in red and blue, respectively.}
  \label{Fig6_TargetXXII_GA_runs}
\end{figure}The molecule contains no rotatable bonds, however it can bend about the S-S axis of the six-membered ring. The energy barrier between its chiral forms is small, leading to the appearance of many structures with planar or near-planar conformations in the computed crystalline energy landscape \cite{2016Reilly, 2016Curtis}. The correct crystal structure of Target XXII was generated by 12 out of 21 groups that participated in category 1 of the most recent blind test\cite{2016Reilly}, and ranked as the most stable structure by 4 groups.

\begin{figure*}[h!]
  \includegraphics{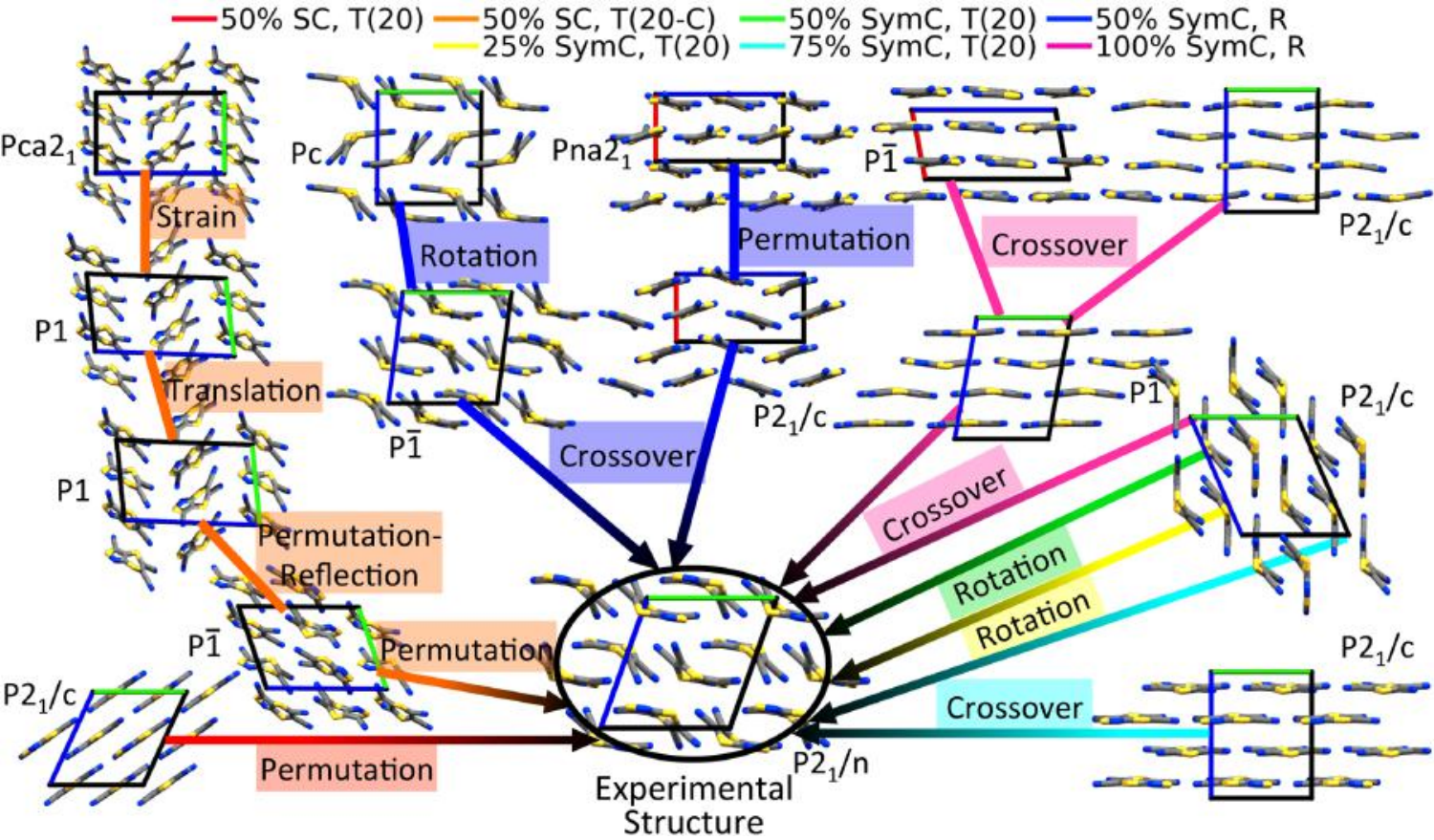}
  \caption{The different evolutionary routes which generated the experimental structure of Target XXII for different runs of the GA. The $\vec{a}$, $\vec{b}$, and $\vec{c}$ crystallographic lattice vectors are displayed in red, green, and blue, respectively.}
  \label{Fig7_TargetXXII_Breeding_Routes}
\end{figure*}
GAtor was run with a variety of GA settings using the same initial pool. In principle, a GA should be run numerous times to determine how a particular group of settings perform. Because GAtor is a first principles algorithm a more practical approach is adopted, where different GA settings are used in several runs and then the structures produced from all runs are combined for postprocessing. For Target XXII, the initial pool contained 100 structures in a variety of space groups. All runs were stopped when the number of structures added to the common population from the GA reached 550 structures. Fig. \ref{Fig6_TargetXXII_GA_runs} shows an analysis of the various GA runs. Here, a GA iteration corresponds to when a single structure has passed all rejection criteria and has been added to the common population. The shorthand notation used for the different GA runs is as follows: standard crossover (SC), symmetric crossover (SymC), tournament selection (T), and roulette wheel selection (R).  The percentage (e.g. 75\%) indicates the crossover probability, with the remaining percentage (e.g. 25\%) indicating mutation probability.  For runs that used tournament selection, the tournament size is shown in parentheses. Cluster-based fitness is denoted by a C after the selection type. Here, Affinity Propagation\cite{2007Frey} clustering was used with the descriptor given by Eq. 3, which promotes the selection  of structures with under-sampled lattice parameters. Although this descriptor is simple, it provides insight into the behavior of cluster-based fitness in the GA and was successful in generating the experimental structure of Target XXII. 

The average energy of the top 20 structures per GA iteration for the different runs is shown in Fig. \ref{Fig6_TargetXXII_GA_runs}, panel (a). The energies shown are relative to the global minimum structure evaluated with PBE+TS and \textit{lower-level} numerical settings. For the seven runs that used energy-based fitness, the average energy of the top 20 structures smoothly converges to within approximately 5 kJ/mol per molecule of the global minimum structure upon GA termination. The runs that used tournament selection had a slightly lower average energy of the top 20 structures over time compared to the runs using roulette wheel selection.  The run that used clustering, depicted in orange, shows a larger average energy than the other runs and a slower, more erratic convergence of the top 20 structures to within 7 kJ/mol per molecule of the global minimum structure upon GA termination. This behavior is not unusual because the cluster-based fitness explicitly promotes under-represented structures in the population, which may have higher energies.

For all runs, the minimum energy structure as a function of GA iteration is shown in Fig. \ref{Fig6_TargetXXII_GA_runs}, panel (b). The energies of the experimental structure and the lowest energy structure in the initial pool are also indicated. The latter happened to correspond to the PBE+TS global minimum structure using \textit{lower-level} numerical settings. We note that the initial pool produced by Genarris is not random, but rather consists of a diverse set of structures pre-screened with a Harris Approximation\cite{1985Harris}, as detailed in Ref. \citenum{2017Li}. All runs generated the experimental structure (located approximately 3.3 kJ/mol per molecule above the global minimum) but at different GA iterations. Most runs located structures lower in energy than the experimental, but only those that used tournament selection and energy-based fitness (shown in red, yellow, green, and cyan) generated the second to the global minimum structure. GA runs that used symmetric crossover, tournament selection, and energy-based fitness (shown in yellow, green, and cyan) found the experimental structure in fewer GA iterations on average than the runs that used energy-based fitness and roulette wheel selection (shown in blue, purple, and pink).

Fig. \ref{Fig7_TargetXXII_Breeding_Routes} depicts different evolutionary routes that generated the experimental structure in selected GA runs. Each route starts from an initial pool structure and details the various breeding operations (followed by local optimization), which ultimately generate the experimental structure. The variety of evolutionary routes and paths highlights the flexibility and randomness of the GA. In particular, the run that utilized the cluster-based fitness function, shown in orange, took a unique path to the experimental structure. A crucial mutation along this route was permutation-reflection, which introduced an inversion center and created a $P\bar{1}$, $Z^\prime$=2 structure. This $P\bar{1}$ structure subsequently underwent permutation followed by local optimization to generate the $P2_1/n$ experimental structure. Overall, the combination of symmetric crossover and mutation was highly effective for Target XXII.
\begin{figure*}[h!]
  \includegraphics{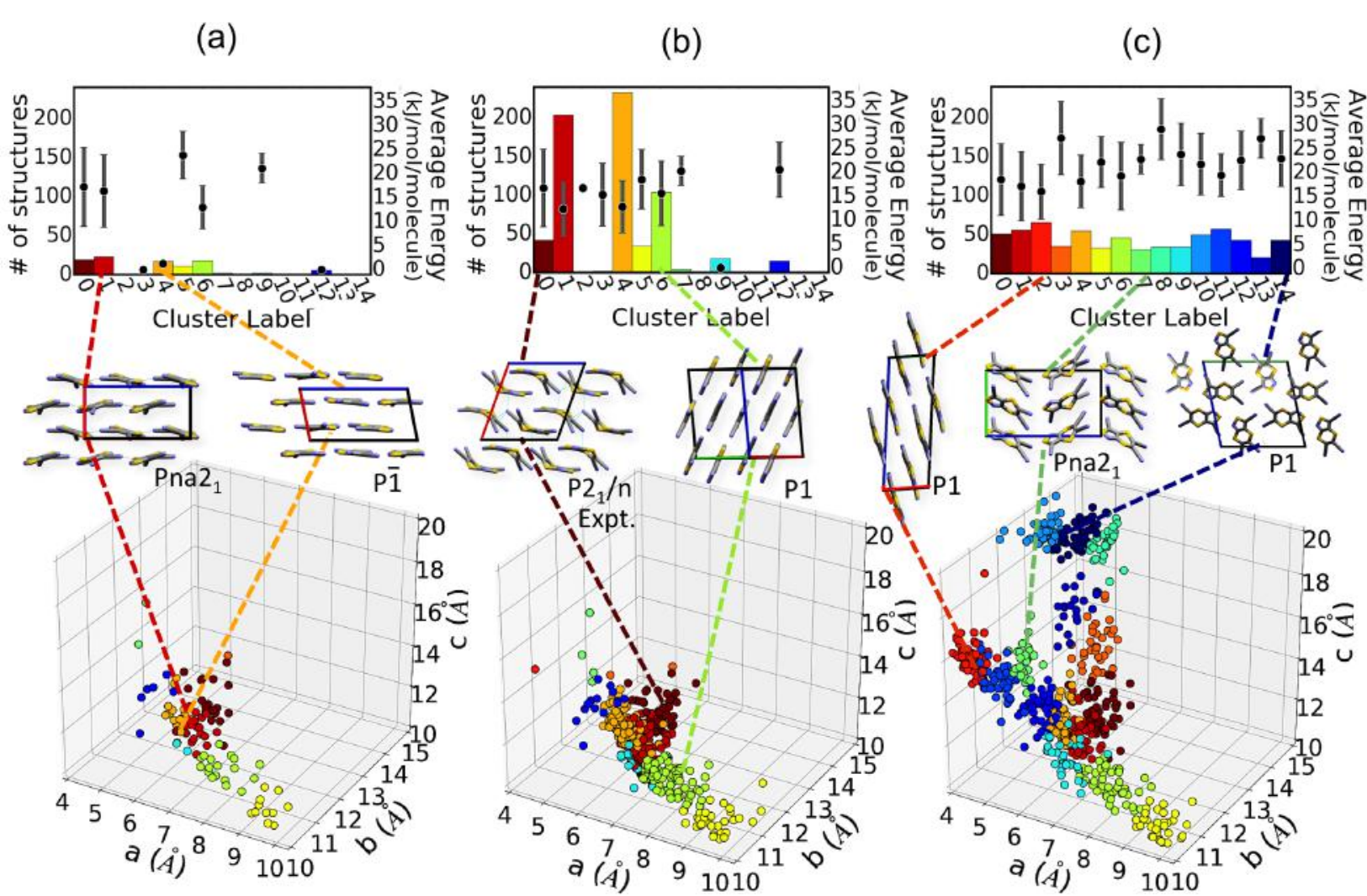}
  \caption{A comparison of the clusters and structural motifs found in (a) the initial pool, (b) the common population evolved using energy-based fitness, and (c) the common population evolved with cluster-based fitness. The average energy for each cluster is plotted using black circles and the standard deviation of energies for each cluster is depicted in grey. For the crystal structures shown, the $\vec{a}$, $\vec{b}$, and $\vec{c}$ crystallographic lattice vectors are displayed in red, green, and blue, respectively.}
  \label{Fig8_TargetXXII_clustering}
\end{figure*}

A detailed comparison between the runs that used tournament selection and 50\% percent standard crossover, with and without cluster-based fitness, is shown in Fig. \ref{Fig8_TargetXXII_clustering}. 
The final structures produced from the cluster-based fitness run, including the initial pool, formed 15 clusters, using Affinity Propagation with the lattice parameter based descriptor and a Euclidean metric. The structures from the run which used energy-based fitness were assigned to one of the 15 clusters from the cluster-based run. Panel (a) depicts the population of the initial pool, while panels (b) and (c) depict the independent evolution of the initial population for the energy and cluster-based fitness runs, respectively. The initial pool contained several low-energy structures with planar or near-planar conformations, which tend to have shorter $a$ parameters than structures with bent conformations, such as the experimental structure. Panel (b) reveals initial pool bias and genetic drift in the run that used energy-based fitness. Initial pool bias is evident from the fact that the GA hardly explores regions not represented in the initial pool. Genetic drift is apparent from the preferential exploration of the clusters labeled 1, 4, and 6, compared to other clusters represented in the initial pool. 
\begin{figure*}[h!]
 \includegraphics{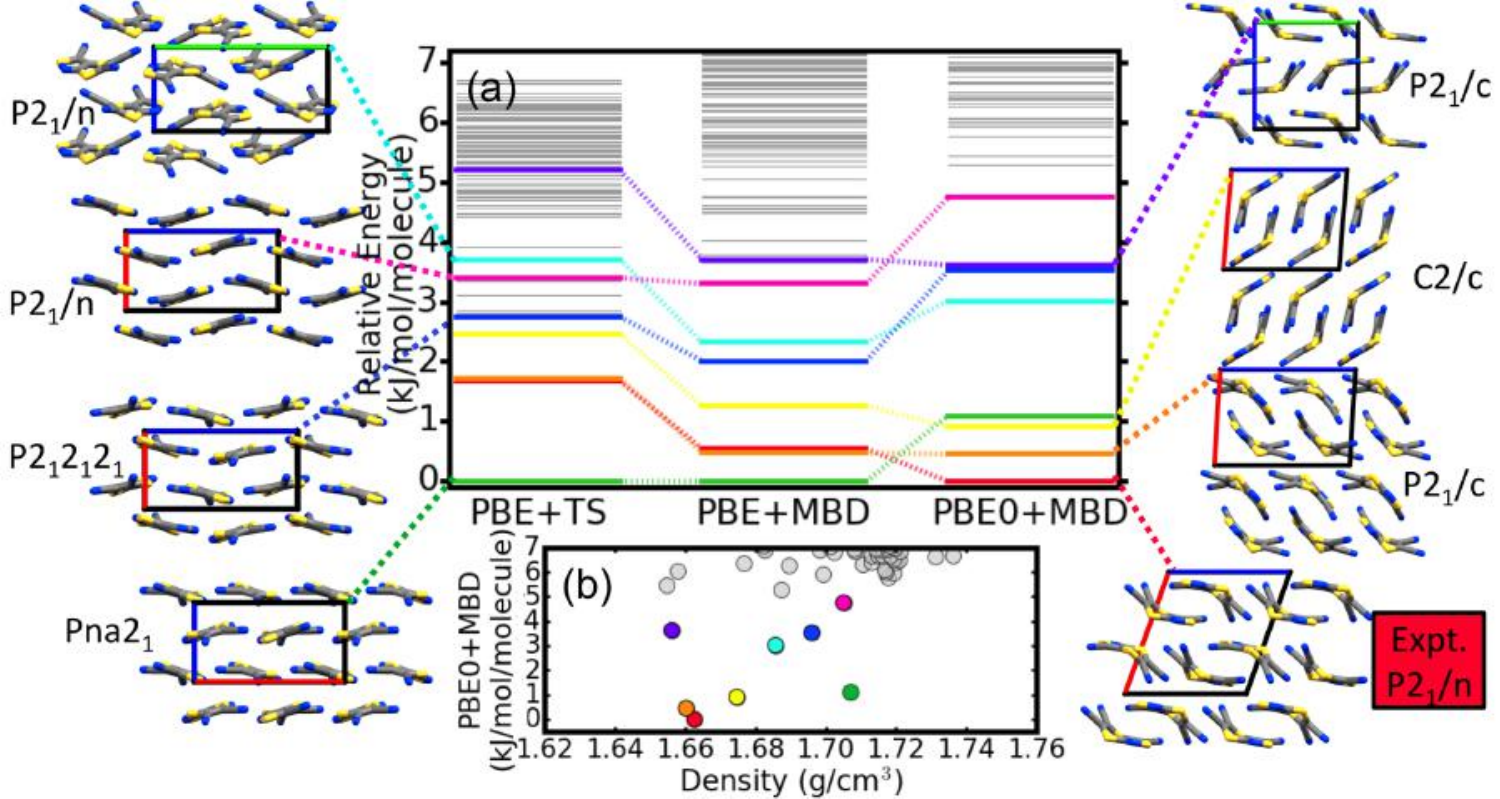}
 \caption{(a) The relative total energies as obtained by different dispersion-inclusive DFT methods and (b) the PBE0+MBD energy versus density of putative crystal structures of Target XXII. The top 8 predicted structures, as ranked by PBE0+MBD, are shown in color. The $\vec{a}$, $\vec{b}$, and $\vec{c}$ crystallographic lattice vectors are displayed in red, green, and blue, respectively.}
  \label{Fig9_TargetXXII_Reranking}
\end{figure*}
These clusters contain layered structures with planar or near-planar conformations, examples of which are shown in panels (a) and (b). Such structures likely correspond to large, shallow basins of the energy landscape that are frequently visited. In addition, these structural motifs are systematically favored by PBE+TS, as discussed in detail in Ref. \citenum{2016Curtis} and below. Cluster 0, which contains structures with a bent conformation, including the experimental structure, is sampled less frequently, possibly because such structures correspond to narrow wells in the potential energy surface that are more difficult to locate. Panel (c) demonstrates that evolutionary niching helps overcome initial pool biases and genetic drift. In this case, a more uniform sampling of the potential energy landscape is achieved. Clusters 1, 4, and 6 have fewer members than in the energy-based run, while cluster 0 has more members. Evidently, for Target XXII, utilizing cluster-based fitness with the lattice parameter descriptor suppressed the over-selection of crystal structures with planar or near-planar conformations. This descriptor was effective for Target XXII because in this case the unit cell shape is correlated with the molecular conformation. Furthermore, several clusters outside the boundaries of the initial pool were only explored with the cluster-based fitness function. These clusters include, for example, structures with more elongated unit cell shapes (a representative structure is shown for cluster 2). This demonstrates that evolutionary niching can correct initial pool biases and explore novel regions of the potential energy surface (this may be particularly useful if the initial pool is not as optimal as the pools produced by Genarris). However, it does so at the price of an increased computational cost, and in this case generates more high-energy structures that may or may not be useful for the purpose of maintaining diversity.

All structures generated were combined into a final set of 200 unique structures evaluated with PBE+TS and \textit{lower-level} numerical settings.  The structures were re-relaxed using PBE+TS with \textit{higher-level} numerical settings and subsequently re-checked for duplicates.  The final 100 PBE+TS structures were then re-ranked with PBE+MBD and PBE0+MBD, as shown in panel (a) of Fig. \ref{Fig9_TargetXXII_Reranking}. The re-ranking of Target XXII structures generated within the sixth CSP blind test has been discussed extensively in Ref. \citenum{2016Curtis}. It has been demonstrated therein that different exchange-correlation functionals and dispersion methods systematically favor specific packing motifs. The experimental structure was ranked as the top structure only by PBE0+MBD. The same trends are observed here. Within the present study, the top 100 $Z$=4 structures are located within relative energy windows of 6.7, 7.5, and 9.6 kJ/mol per molecule using PBE+TS, PBE+MBD, and PBE0+MBD, respectively. The number of structures generated in these intervals shows significant improvement compared to our submission to the sixth blind test. In particular, an important low-energy structure (ranked as \#3 by PBE0+MBD) was located in the present study in addition to the experimental structure. These improvements may be attributed to a number of factors including updated crossover, mutation, and similarity checks, as well as the use of a more diverse and comprehensive initial pool as generated by Genarris \cite{2017Li}. Panel (b) of Fig. \ref{Fig9_TargetXXII_Reranking} shows the PBE0+MBD energy versus density for the structures. Structures with bent molecular conformations, including the experimental structure, have lower densities than structures with planar or near-planar conformations.

\subsection{Target II}
\begin{figure}[h!]
  \includegraphics{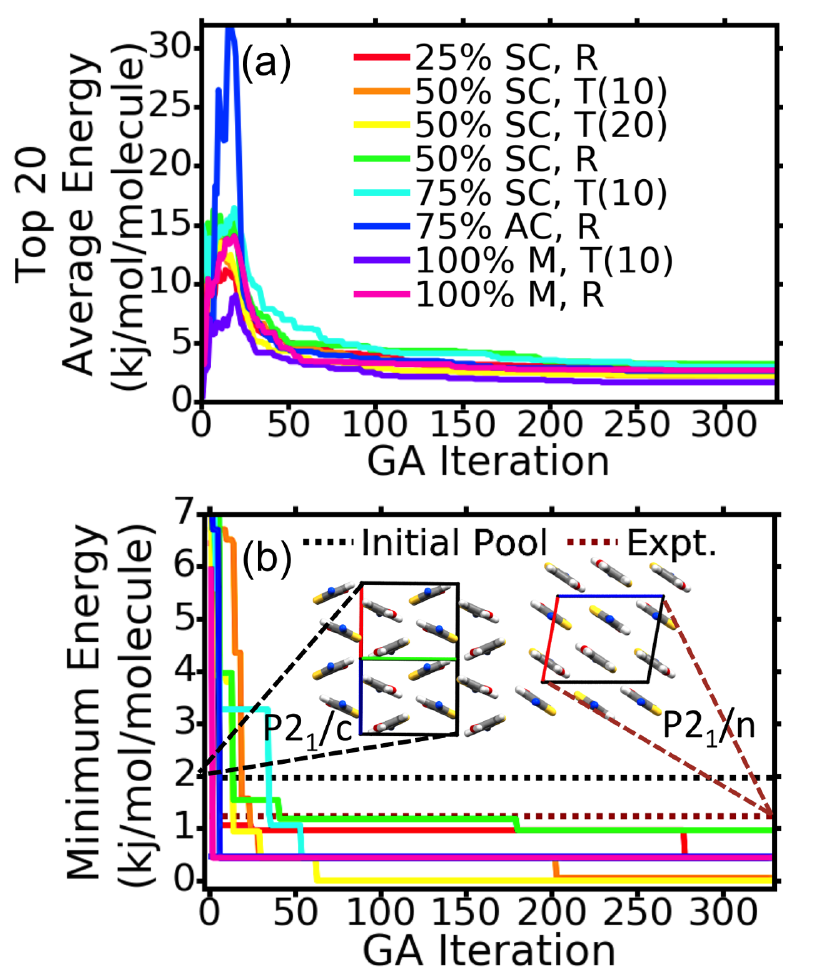}
  \caption{(a) The average energy of the top 20 Target II structures as a function of GA iteration and (b) the global minimum structure generated as a function of GA iteration, shown for different GA runs. S, N, O, C, and H atoms are colored in yellow, blue, red, grey, and white, respectively. The $\vec{a}$, $\vec{b}$, and $\vec{c}$ crystallographic lattice vectors are displayed in red, green, and blue, respectively.}
  \label{Fig10_TargetII_GA_runs}
\end{figure}

Target II (C$_5$H$_3$NOS) was selected from the second blind test \cite{2002Motherwell,1999Blake}. At the time, no participating groups used \textit{ab initio} methods for the structure prediction of this molecule, and only one group submitted the correct experimental structure, ranking it as their second most thermodynamically stable structure. 
Fig. \ref{Fig10_TargetII_GA_runs} shows an analysis of the different GA runs that successfully generated the experimental crystal structure of Target II. The initial pool contained 45 structures. Each run was stopped when the number of additions to the common pool reached 350. 
\begin{figure*}[h!]
  \makebox[\textwidth][c]
      {\includegraphics{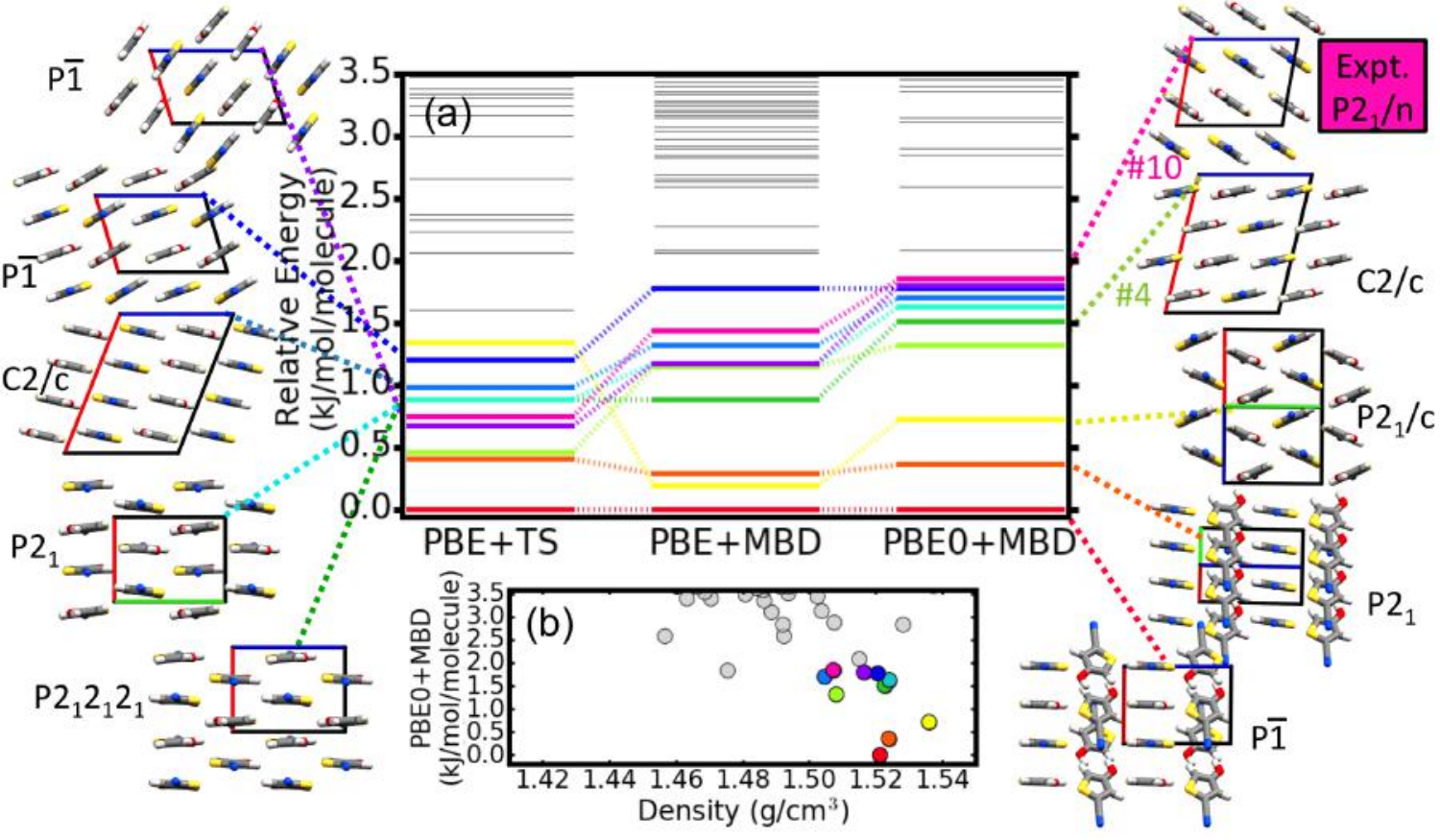}}
 \caption{(a) The relative total energies as obtained by different dispersion-inclusive DFT methods and (b) the PBE0+MBD energy versus density of putative crystal structures of Target II. The top 10 predicted structures, as ranked by PBE0+MBD, are shown in color. The $\vec{a}$, $\vec{b}$, and $\vec{c}$ crystallographic lattice vectors are displayed in red, green, and blue, respectively.}
  \label{Fig11_TargetII_Reranking}
\end{figure*}
The average energy of the top 20 structures as a function of GA iteration is shown in panel (a) of Fig. \ref{Fig10_TargetII_GA_runs}. All energies shown are relative to the energy of the global minimum structure as ranked by PBE+TS with the \textit{lower-level} numerical settings used within the GA. All runs converged the top 20 structures to within 4 kJ/mol per molecule when the GA was terminated. The run that used 100\% pure mutation (denoted by 100\% M) with tournament selection, shown in purple, consistently exhibited the lowest average energy of the top 20 structures. In panel (b), the minimum energy structure added by the GA as a function of GA iteration is shown along with the lowest-energy structure from the initial population and the experimental structure.  All runs generated the experimental structure, as well as at least one other structure lower in energy. Two runs, shown in orange and yellow, generated the most structures with lower energies than the experimental. Strain mutations were particularly effectively at generating new low-energy structures for this target.

All structures produced by the different GA runs were combined into a final set of 200 non-duplicate structures as evaluated with PBE+TS and \textit{lower-level} numerical settings.  The structures were re-relaxed with PBE+TS and \textit{higher-level} numerical settings and subsequently re-checked for duplicates.  The final 100 PBE+TS structures were then re-ranked with PBE+MBD and PBE0+MBD, as shown in panel (a) of Fig. \ref{Fig11_TargetII_Reranking}. The top 10 structures as ranked by PBE0+MBD are highlighted in color. The top 100 structures are found in relative energy windows of 5.3, 5.5, and 6.1 kJ/mol per molecule using PBE+TS, PBE+MBD, and PBE0+MBD, respectively. Interestingly, the experimental structure becomes less stable with increasingly accurate DFT methods and is ranked as \#10 with PBE0+MBD. Structures ranked as \#4-\#10 with PBE0+MBD display layered packing motifs in several different space groups, within an energy window of approximately 0.6 kJ/mol per molecule. The layered motif of Target II is characterized by hydrogen-bonds that form 1D chains between the hydroxyl group of one molecule and the nitrile group of another (O$-$H$\cdot\cdot\cdot$N) that are stacked on top of one another as shown in Fig. S1 in the supporting information. The prediction of nearly energetically degenerate crystal structures consisting of the same sheet stacked in different ways is a common phenomena\cite{2013Price,2012Braun,2016Braun}. While the structures ranked \#4-\#10 are determined as distinct lattice energy minima, they likely converge to a lower number of minima on the free energy surface\cite{2013Price, 2017Whittleton}. 

The structure ranked as \#3 by PBE0+MBD (shown in yellow) was not reported by any participating group during the second blind test and has the highest computed density of the low-energy structures, as shown in panel (b) of Fig. \ref{Fig11_TargetII_Reranking}. This structure contains the same 1D hydrogen-bonded patterns as the experimental structure, but with zig-zag stacking.  Ref. \citenum{2011Chan} later performed an additional CSP study on Target II using a tailor-made force field\cite{2008Neumann} within the GRACE software. This methodology has been highly successful at CSP and predicted all five targets in the most recent blind test\cite{2016Reilly}. Searching structures with $Z^\prime$=1, this study predicted the \#3 PBE0+MBD zig-zag structure for the first time, ranking it as the global minimum structure when re-ranked using DFT with a pairwise dispersion correction.
\cite{2011Grimme}. Furthermore, it was shown that this form became more stable with increasing pressure, suggesting it could be an unobserved high-pressure polymorph of Target II. Our \#2 $P2_1$ PBE0+MBD structure with a scaffold packing motif was also discussed in Ref. \citenum{2011Chan} and ranked as \#3. Ref. \citenum{2017Whittleton} computed the relative stability of the $P2_1$ scaffold structure and the experimental structure using the B86bPBE density functional\cite{1986Becke,1996Perdew} combined with the exchange-hole dipole moment (XDM)\cite{2007Becke,2012Otero_2} dispersion model and found the $P2_1$ scaffold structure to be more stable. When a quasi-harmonic thermal correction was further included, the experimental structure was ranked as the more stable structure. 

The $P\bar{1}$, $Z^\prime$=2 structure with a scaffold packing motif ranked as the global minimum by all three DFT methods has not been reported in any previous CSP studies of Target II. It has a higher computed density than the experimental form, as shown in panel (b) of Fig. \ref{Fig11_TargetII_Reranking}. A discussion comparing the packing motifs of the experimental structure and the \#1 PBE0+MBD scaffold structure is provided in the supporting information. The \#1 PBE0+MBD scaffold structure would not have been found without GAtor's ability to generate crystal structures with Z$^\prime>$1 through the various crossover and mutation operators. As emphasized in Ref.\citenum{2015Steed}, stable crystal structures are formed when intermolecular interactions are optimized through close packing. While these requirements favor highly symmetric structures, symmetry can be sacrificed in favor of forming particularly stabilizing intermolecular interactions\cite{2015Steed,2010Vande,2016Taylor}. Future investigations incorporating finite temperature and pressure effects will add further insight into the relative stability of the \#1 PBE0+MBD scaffold structure and the other predicted low-energy structures, including the experimental.

\subsection{Target XIII}

Target XIII (C$_6$H$_2$Br$_2$ClF) was selected from the fourth blind test \cite{2009Day}, in which it was categorized as a rigid molecule containing challenging elements for modeling methods. Target XIII contains three different halogens, allowing for a variety of halogen bonds. Many common electronic structure theory  methods do not accurately capture halogen bonds because they require a precise treatment of both electrostatic and dispersion interactions\cite{2008Riley,2016Cavallo,2013Kozuch,2012Rezac,2014Otero-de-la-Roza}. During the fourth blind test, the correct experimental structure was successfully predicted and ranked as \#1 by 4/14 groups. The methodology used in one of the successful submissions is further detailed in Ref. \citenum{2008Misquitta}.
\begin{figure}[h]
  \includegraphics{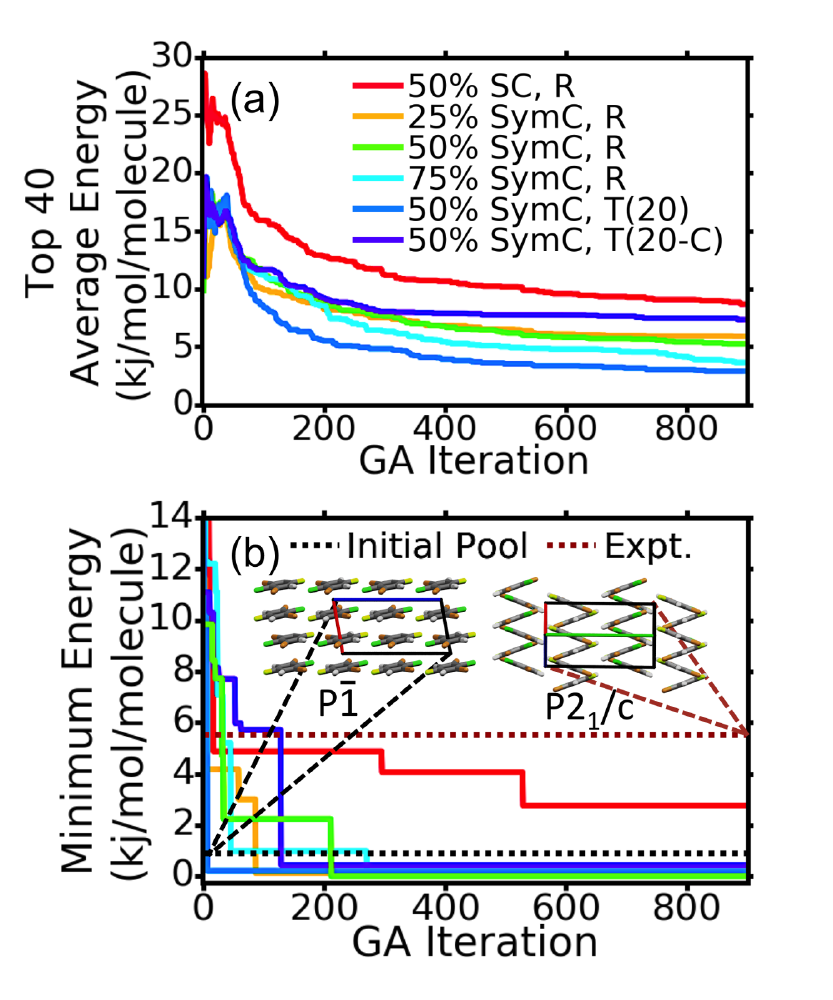}
  \caption{(a) The average energy of the top 40 Target XIII structures and (b) the global minimum structure produced as a function of GA iteration for different GA runs. C, H, Br, Cl, and F atoms are colored in grey, white, brown, green, and yellow, respectively. The $\vec{a}$, $\vec{b}$, and $\vec{c}$ crystallographic lattice vectors are displayed in red, green, and blue, respectively.}
  \label{Fig12_TargetXIII_GA_runs}
\end{figure}
\begin{figure*}[h!]
    \makebox[\linewidth][c]{\includegraphics{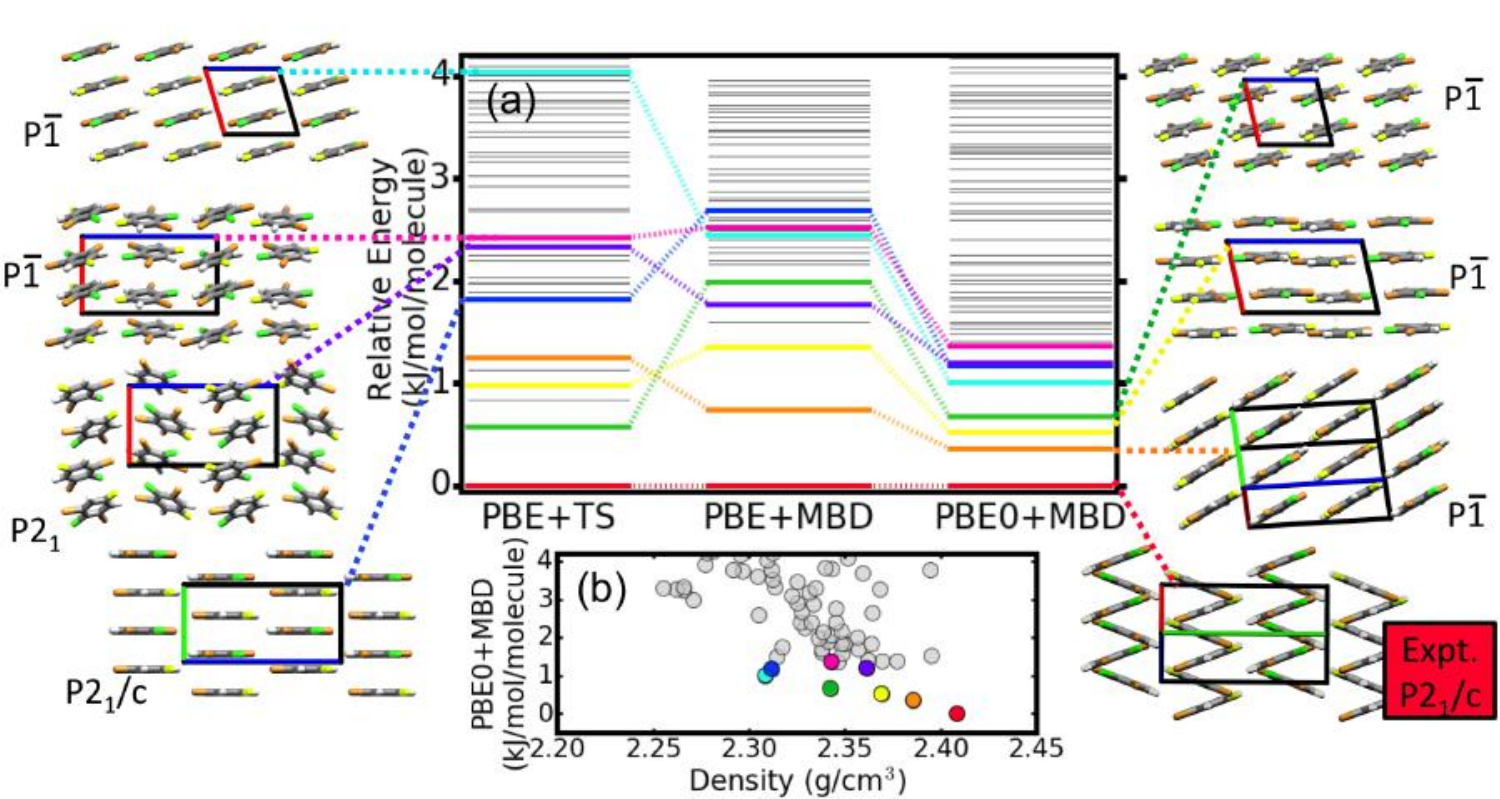}}
 \caption{(a) The relative total energies as obtained by different dispersion-inclusive DFT methods and (b) the PBE0+MBD energy versus density of putative crystal structures of Target XIII. The top 8 predicted structures, as ranked by PBE0+MBD, are shown in color. The $\vec{a}$, $\vec{b}$, and $\vec{c}$ crystallographic lattice vectors are displayed in red, green, and blue, respectively.}
  \label{Fig13_TargetXIII_Reranking}
\end{figure*}

Indeed, predicting the correct crystal structure of Target XIII proved challenging. The various crossover, mutation, and selection settings used in different GA runs of Target XIII are shown Fig. \ref{Fig12_TargetXIII_GA_runs}. The initial pool for all runs contained 48 structures. The various GA runs were stopped after 1400 iterations, the first 900 of which are shown. Of the six runs attempted, only one run (50\% SymC, R), colored in green, generated the experimental structure, although all runs found crystal structures lower in energy than the experimental using PBE+TS and \textit{lower-level} numerical settings.  Panel (a) shows the average energy of the top 40 structures as a function of GA iteration, relative to the global minimum energy with PBE+TS and \textit{lower-level} settings. The run that used standard crossover (50\% SC, R), colored in red, consistently had the highest average energy, even higher than the run that used cluster-based fitness with Affinity Propagation and the lattice parameter based descriptor, shown in indigo. Panel (b) shows the minimum energy structure as a function of GA iteration. All runs converged the top structure to within 1 kJ/mol per molecule within 300 iterations, except for the run that used standard crossover. For this target symmetric crossover was essential in producing low-energy structures.

All structures generated were combined into a final set of 200 unique structures evaluated with PBE+TS and \textit{lower-level} numerical settings. The top 150 structures were re-relaxed with PBE+TS with \textit{higher-level} numerical settings and subsequently re-checked for duplicates. The final top 90 structures as ranked by PBE+TS and \textit{higher-level} settings, were then re-ranked with PBE+MBD and PBE0+MBD. The top 90 structures are located within relative energy windows of 6.8, 7.8, and 6.5 kJ/molecule per molecule when ranked by PBE+TS, PBE+MBD, and PBE0+MBD, respectively. Panel (a) of Fig. \ref{Fig13_TargetXIII_Reranking} shows 
the ranking of the structures found within a window of 4.2 kJ/mol per molecule of the global minimum.  The top 8 crystal structures as ranked by PBE0+MBD are highlighted in color.  After local optimization with PBE+TS and \textit{higher-level} numerical settings, the experimental structure is ranked as \#1. It is consistently predicted as the most stable crystal structure by PBE+MBD and PBE0+MBD. Focusing on the top 8 crystal structures as ranked by PBE0+MBD, only the experimental structure contains a zig-zag packing motif. Additionally, 4/8 of the top structures have $Z^\prime$=2. Panel (b) of Fig. \ref{Fig13_TargetXIII_Reranking} shows the PBE0+MBD energy versus density of the top structures. This reveals the experimental structure with the zig-zag motif has the highest density.  For the experimental structure, close bromine-bromine contacts are found perpendicular to the zig-zag stacking direction, while 7/8 of the other top structures generated show $\pi$-stacking and/or close halogen bonds that stabilize the stacking of the layers. 

Although many low-energy structures were generated, 5/6 of the GA runs did not successfully locate the experimental structure. This may be attributed to two primary factors. First, it is possible that the \textit{lower-level} numerical settings used to save computational time in the GA search, were not sufficiently accurate for this halogenated molecule. When using PBE+TS and \textit{lower-level} numerical settings, the experimental structure was nearly 6 kJ/mol per molecule higher than the global minimum, and ranked as \#39 when all structures generated from the different GA runs were combined. When these structures were postprocessed with PBE+TS and \textit{higher-level} numerical settings, the experimental structure was ranked as \#1. As lower energy structures have a higher probability of being selected, this could have systematically biased the searches. This highlights the complications that may arise when using a hierarchical approach. Second, while most low-energy structures of Target XIII have a layered packing motif, the experimental structure has a unique zig-zag packing motif and an oblong unit cell. Such oblong unit cells were rarely generated in the search. In fact, even the run that used cluster-based fitness with the lattice parameter descriptor failed to locate the experimental structure. Although candidate child structures with similar lattices to the experimental were frequently generated in this run, they were subsequently rejected by the geometric and energetic constraints before local optimization. This suggests that the experimental structure is located in a narrow well in the potential energy surface, while layered structures exist in wider, more-accessible basins. When studying halogen-bonded systems in the future, it may be beneficial to use cluster-based fitness with a descriptor based on halogen-halogen or hydrogen-halogen intermolecular contacts.

\subsection{Target I}
Target I (C$_6$H$_6$O) was selected from the second blind test \cite{2002Motherwell,1999Blake}. It has two reported polymorphs, a stable form, which crystallizes in $P2_1/c$ with $Z$=$4$, and a metastable form which crystallizes in $Pbca$ with $Z$=$8$. At the time of the second blind test, no participating groups submitted the more stable $Z$=$4$ form. 4/11 groups submitted the metastable $Z$=$8$ form, with 3/4 groups ranking it as the most stable structure.
\begin{figure}[h!]
  \includegraphics{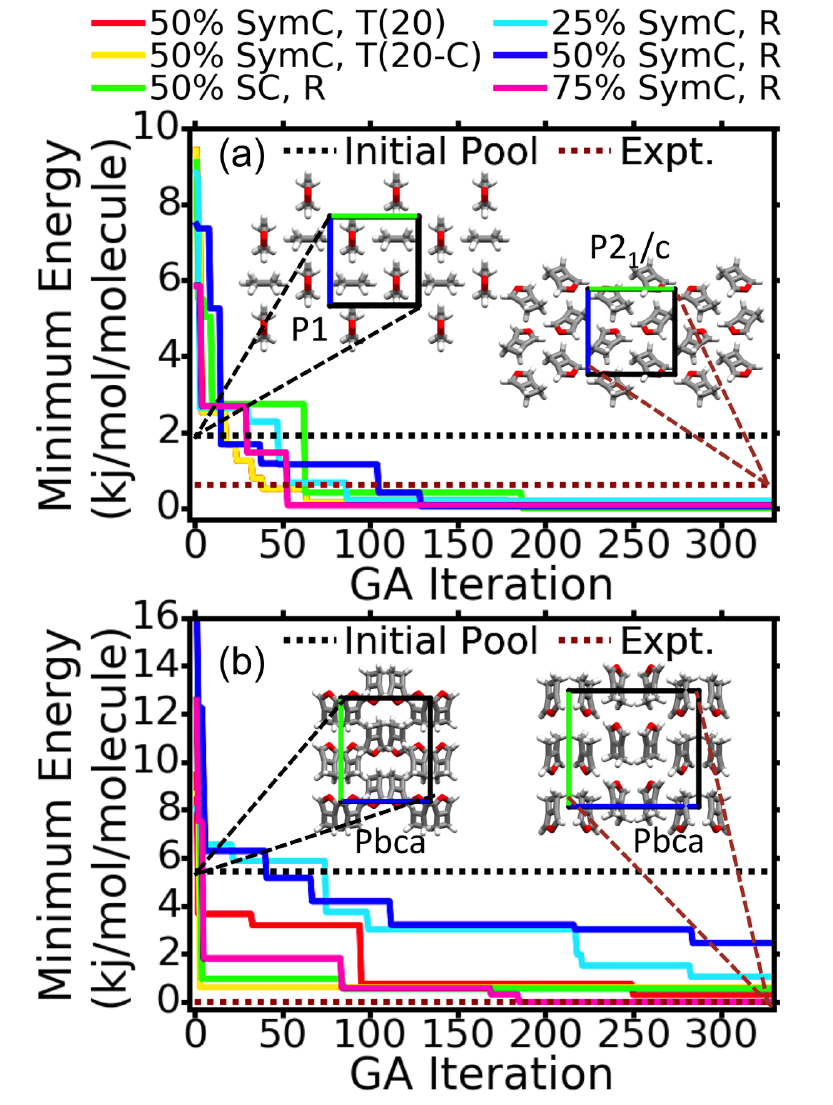}
  \caption{The global minimum structure produced by the GA runs as a function of GA iteration for runs that used (a) $Z$=4 and (b) $Z$=8. C, H, and O atoms are colored in grey, white, and red, respectively.  The structures shown are projected along the $\vec{a}$ lattice vector and the $\vec{b}$, and $\vec{c}$ lattice vectors are highlighted in green and blue, respectively.}
  \label{Fig14_TargetI_GA_runs}
\end{figure}
\begin{figure*}[h!]
\includegraphics{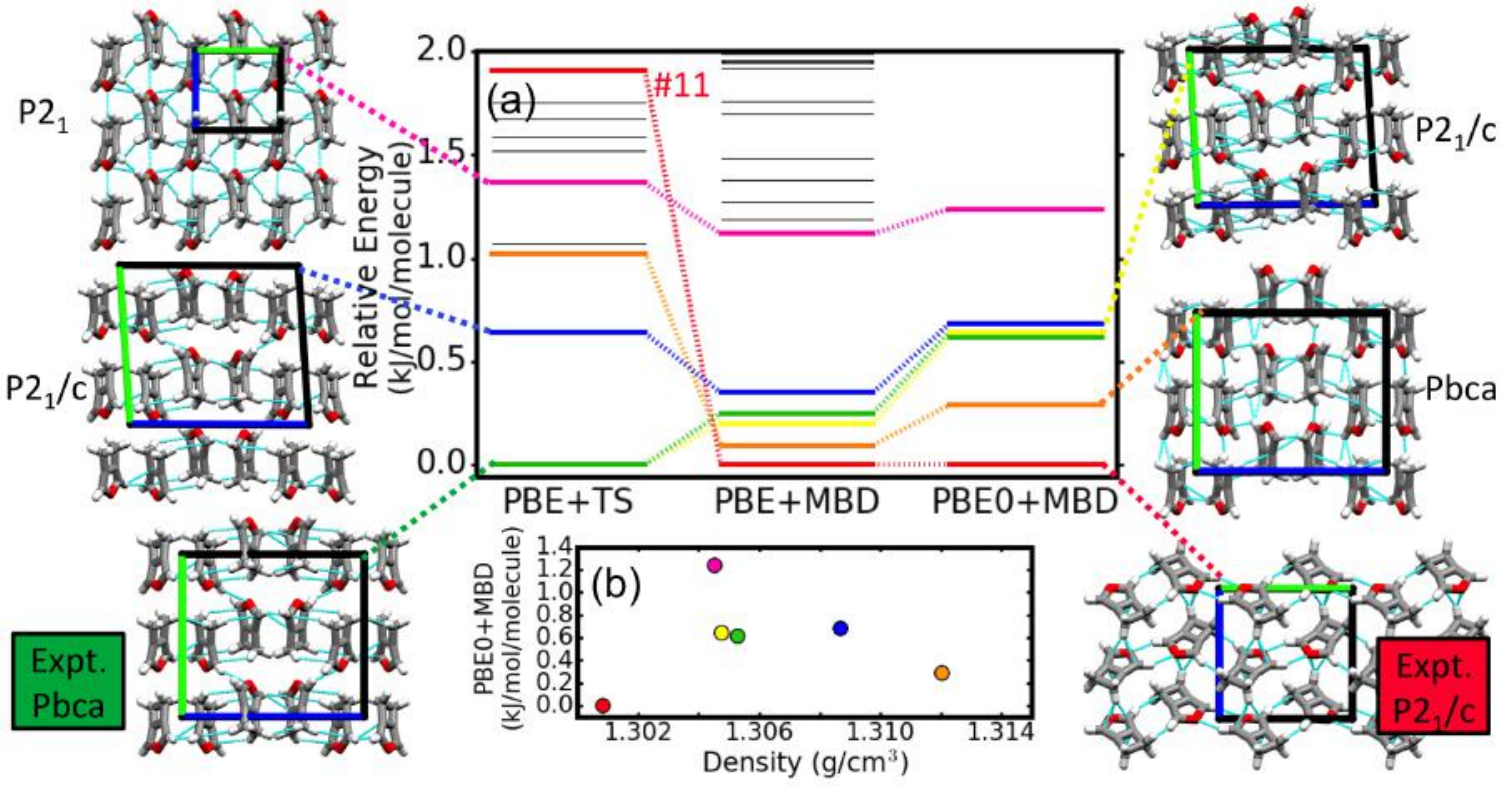}
 \caption{(a) The relative total energies as obtained by different dispersion-inclusive DFT methods and (b) the PBE0+MBD energy versus density of selected crystal structures of Target I. The top 6 predicted structures, as ranked by PBE+MBD, are highlighted in color. Intermolecular contacts less than the sum of vdW radii are shown in cyan.  The structures shown are projected along the $\vec{a}$ lattice vector and the $\vec{b}$, and $\vec{c}$ lattice vectors are highlighted in green and blue, respectively.}
  \label{Fig15_TargetI_Reranking}
\end{figure*}
For Target I, independent GA searches were conducted starting from initial pools with $Z$=4 and $Z$=8. These contained 45 and 96 structures, respectively. The GA runs were stopped when the number of additions to the common pool reached 650 and 350, respectively. During evolution, the $Z$=4 runs also generated structures with $Z$=2, and the $Z$=8 runs generated structures with $Z$=4 and $Z$=2. The minimum energy as a function of GA iteration, relative to the global minimum using PBE+TS with \textit{lower-level} numerical settings, is shown in Fig. \ref{Fig14_TargetI_GA_runs}, panels (a) and (b), for the $Z$=4 and $Z$=8 runs, respectively. For the $Z$=4 runs, the convergence behavior of the minimum energy structure was similar for all settings tested, including the run that used lattice parameter based clustering, shown in orange. All runs located structures lower in energy than the $Z$=4 experimental polymorph at this level of theory. For the $Z$=8 runs, the runs that used 25\% and 50\% symmetric crossover with roulette wheel selection were slower to converge, and did not locate the $Z$=8 polymorph when the GA was stopped.

All structures produced by the $Z$=$4$ and $Z$=$8$ GA runs were combined into a final set of 200 unique structures, as evaluated with PBE+TS and \textit{lower-level} numerical settings. Supercells were allowed in the pymatgen duplicate check. The final top 100 structures were re-relaxed using PBE+TS with higher-level settings and subsequently re-ranked using PBE+MBD. The structures located within 2 kJ/mol per molecule of the global minimum are shown in panel (a) of Fig. \ref{Fig15_TargetI_Reranking}. The top 6 structures as ranked by PBE+MBD were also re-ranked using PBE0+MBD and are highlighted in color.  Of these top 6 structures, 4/6 display similar packing motifs to the metastable $Pbca$ polymorph, shown in green with co-facial dimers oriented in opposite directions, stacked in slightly different ways. To highlight structural differences, intermolecular close-contacts are displayed in cyan. 

The metastable $Z$=$8$ $Pbca$ polymorph, shown in green, is ranked as \#1 with PBE+TS, \#4 with PBE+MBD, and \#3 when re-ranked with PBE0+MBD. With all energy methods this polymorph is determined to be practically energetically degenerate with the putative $Z$=8 $P2_1/c$ structure, shown in yellow. However, the $Z$=8 $P2_1/c$ structure has $Z^\prime$=2 and a slightly different lattice from the metastable polymorph, and hence was determined to be a unique lattice energy minima. The experimental $P2_1/c$ polymorph with $Z$=4, highlighted in red, is ranked as \#11 with PBE+TS, but \#1 with PBE+MBD and PBE0+MBD. There is no significant re-ranking between PBE+MBD and PBE0+MBD for the structures considered. The relative energy differences between these structures increased when re-ranked by PBE0+MBD, as compared to PBE+MBD. Panel (b) of Fig. \ref{Fig15_TargetI_Reranking} shows the PBE0+MBD energy versus density of the highlighted structures. The six structures have very similar densities, but the most stable $P2_1/c$ experimental structure has the lowest density.

Several computational studies conducted after the second blind test\cite{2001Mooij,2004Day,2009Asmadi} consistently ranked the $Z$=8 $Pbca$ polymorph as the most stable form. However, attempts at its recrystallization only lead to the stable $Z$=4 $P2_1/c$ form. Ref. \citenum{2015Nyman} suggests that the $Z$=8 $Pbca$ structure is located on a saddle point of the potential energy surface and that symmetry breaking produces a stable $Z^\prime$=2 structure. This could be the $Z^\prime$=2 $P2_1/c$ structure, colored in yellow and ranked as \#4 with PBE0+MBD, as discussed above. It should be noted, however, that the nature of the potential energy landscape, including whether certain structures are determined as minima or saddle points, may depend strongly on the energy method used \cite{2014Wales, 2016Carr}. In the present study, PBE+MBD and PBE0+MBD rank the experimental $Z$=4 $P2_1/c$ structure as the most stable polymorph. This highlights the importance of accounting for many-body dispersion interactions and long-range screening effects in the MBD method. Ref. \citenum{2017Whittleton} also computed the $P2_1/c$ experimental structure as more stable than the $Pbca$ form using B86bPBE-XDM. 

\section{Conclusion and Best Practices}
We have introduced GAtor, a first principles genetic algorithm for molecular crystal structure prediction. GAtor currently interfaces with FHI-aims and is optimized for HPC environments. The code offers a variety of features that enable the user to customize the GA search settings, including energy-based and cluster-based fitness (evolutionary niching), roulette wheel and tournament selection, symmetric and standard crossover, different mutation schemes, and various tunable parameters related to energy cutoffs, similarity checks, and geometric constraints. GAtor's crossover and mutation operators, specifically tailored for molecular crystals, provide a balance between exploration and exploitation. These operators enable the generation and exploration of high $Z^\prime$ structures.

GAtor was applied to predict the structures of a chemically diverse set of four past blind test targets. The known structures of all four targets were successfully predicted, as well as several additional low-energy structures. Different GA settings were found to be more effective for different targets. Target XXII contains only C, N, and S atoms and has a small energy barrier between its two enantiomers, related by a bending degree of freedom. For this target, symmetric crossover and tournament selection were particularly effective. Evolutionary niching with respect to a descriptor based on lattice parameters uniformly explored the potential energy surface, including regions outside the initial pool, and suppressed the oversampling of structures with a planar molecular conformation (genetic drift). Target II forms various hydrogen-bonds. Its known experimental structure was located with a variety of GA settings, including runs that purely used mutations. For this molecule, standard crossover was more effective than symmetric crossover. Target XIII contains several halogens (Br, Cl, F), which make it challenging due to the presence of halogen bonds. In addition, the experimental structure comprises a ziz-zag packing motif unlike the layered packing motifs found in most of the low-energy structures in the population. This may explain why the experimental structure was generated only once. For Target XIII, symmetric crossover was critical for the production of low-energy structures. Target I forms mainly weak C$\cdot\cdot\cdot$H and C$-$H$\cdot\cdot\cdot$O interactions. It has two known polymorphs with $Z$=4 and $Z$=8, the latter of which is a less stable ``disappearing polymorph". All GA settings tested were found to be equally effective in generating important low-energy $Z$=4 structures. For the $Z$=8 structure, the combination of 25\% or 50\% symmetric crossover with roulette wheel selection was less effective.

Low-energy structures found in different GA runs were grouped together, re-relaxed, and re-ranked with increasingly accurate dispersion-inclusive DFT methods: PBE+TS, PBE+MBD, and PBE0+MBD. For Target XIII, all three methods ranked the experimental structure as \#1. For Target I, PBE+MBD and PBE0+MBD correctly ranked the $Z$=4 polymorph as \#1 and the $Z$=8 polymorph as less stable, at \#4 and \#3, respectively, and very close in energy to a structure with $Z^\prime$=2 and a similar packing motif. The MBD method was instrumental in obtaining the correct ordering of the two known polymorphs of Target I based solely on lattice energy without considering vibrational and thermal contributions. For Target XXII, only PBE0+MBD ranked the experimental structure as \#1. Target II is an exception because the relative energy of its experimental structure increases, rather than decreases, with increasing accuracy. It is ranked as \#10 with PBE0+MBD. The structures ranked \#4-\#9 exhibit a variety of layered packing motifs, similar to the experimental structure.  The structure consistently ranked as \#1 with all three methods was predicted for the first time using GAtor. It is a $Z^\prime$=2 structure with $P\bar{1}$ symmetry and a scaffold packing motif, whose lattice energy is 1.8 kJ/mol per molecule lower than the known experimental form. The \#2 structure, which also has a scaffold packing motif, and the \#3 structure with a zig-zag packing motif have been previously reported by others. Several of the low-lying putative structures of Target II have higher densities than the observed structure, therefore it may be possible to crystallize them under high pressure conditions. This may motivate further experimental investigations of Target II. Further computational studies considering finite temperature and pressure effects may provide additional insight into the relative stability of the putative low-energy structures identified here and the possibility of growing them experimentally.

Several best practices for the usage of GAtor have emerged from the results reported here. First, because the GA exhaustively explores regions of the configuration space represented in the initial pool (unless evolutionary niching is used), it is recommended to start GAtor from a carefully crafted initial pool, containing a diverse set of structures in all space groups appropriate for the molecule. Such an initial pool may be generated by Genarris \cite{2017Li} or by other means. Second, rather than running GAtor with predetermined settings for a large number of iterations, we recommend running GAtor with several different settings for a smaller number of iterations, and then combining the structures found in all searches for post-processing. As each system is unique and it is difficult to know \textit{a priori} which settings will be the most effective, running the GA with different settings increases the likelihood of success. Third, it is recommended to use evolutionary niching in at least one of the runs. Overall, the goal is to locate all the low-lying minima including those found in disconnected, hard to reach regions of the potential energy surface. For this reason, cluster-based fitness may be a useful tool for uniformly sampling the potential energy landscape and for overcoming initial pool biases and selection biases (genetic drift). In the future, we plan to implement increasingly sophisticated capabilities in GAtor to treat more complex systems. We expect GAtor to be a useful tool for the computational chemistry, materials science, and condensed matter communities.

\begin{acknowledgement}
Work at CMU was funded by the National Science Foundation (NSF) Division of Materials Research through grant DMR-1554428. An award of computer time was provided by the Innovative and Novel Computational Impact on Theory and Experiment (INCITE) program. This research used resources of the Argonne Leadership Computing Facility, which is a DOE Office of Science User Facility supported under Contract DE-AC02-06CH11357.
\end{acknowledgement}


\begin{suppinfo}
The supporting information provides a comparison between the experimental structures predicted by GAtor and the published experimental forms, including RMS differences computed with the Mercury\cite{2008Macrae} software. The distribution of space groups for the different initial pools used is also included. For each target, CIF files and total energies of the structures used for re-ranking are provided.
 \end{suppinfo}
%
%

\bibliography{GAtor_bibliography_abbrevU}

\end{document}